\documentclass[%
preprint,
 amsmath,amssymb,
 prd,
]{revtex4-1}

\usepackage{graphicx}
\usepackage{dcolumn}
\usepackage{bm}
\def\comment#1{}
\usepackage{color}

\begin{document}

\preprint{APS/123-QED}

\title{Dynamics of extended bodies with spin-induced quadrupole in Kerr spacetime: generic orbits}

\author{Wen-Biao Han $^1$ }
 \email{wbhan@shao.ac.cn}
 \author{ Ran Cheng $^{1,2}$}
\affiliation{%
 1. Shanghai Astronomical Observatory, Shanghai, 200030, China   \\
 2. University of Chinese Academy of Sciences,  Beijing 100049, China}





\begin{abstract}
We discuss motions of extended bodies in Kerr spacetime by using Mathisson-Papapetrou-Dixon equations.  We firstly solve the conditions for circular orbits, and calculate the orbital frequency shift due to the mass quadrupoles. The results show that we need not consider the spin-induced quadrupoles in extreme-mass-ratio inspirals for spatial gravitational wave detectors. We quantitatively investigate the temporal variation of rotational velocity of the extended body due to the coupling of quadrupole and background gravitational field. For generic orbits, we numerically integrate the Mathisson-Papapetrou-Dixon equations for evolving the motion of an extended body orbiting a Kerr black hole. By comparing with the monopole-dipole approximation, we reveal the influences of quadrupole moments of extended bodies on the orbital motion and chaotic dynamics of extreme-mass-ratio systems. We do not find any chaotic orbits for the extended bodies with physical spins and spin-induced quadrupoles. Possible implications for gravitational wave detection and pulsar timing observation are outlined.  
\end{abstract}

\pacs{Valid PACS appear here}
\maketitle
\section{Introduction}
The geodesic motion of test particles orbiting central gravitational bodies has been studied one hundred years since the birth of general relativity. The test particle approximation is the most simplified one: omitting spin, structure and gravitational self-force of the small bodies. The geodesic motion of test particles is quite important in revealing the space-time property of central black holes, and also useful in celestial mechanics and astrophysics if the mass $m$ of the small body is very small comparing with the one ($M$) of central black hole. For example, by observing the S stars near the central black hole of our Galaxy, one can test general relativity and determine the physical parameters of the black hole \cite{han14}. The extreme-mass-ratio inspirals (EMRIs)  (compact stars with stellar mass like as white dwarfs, neutron stars or black holes orbiting supermassive black holes) are very important sources for the spatial gravitational wave (GW) detectors such as eLISA\cite{elisa}, Taiji\cite{taiji} and Tianqin\cite{tianqin}.  However, if the mass-ratio $m/M \sim 10^{-5}$,  a few literature have shown that one should at least consider the dipole approximation (spin) to replace the test particle one \cite{tanaka96,han10, gair11,harms16,hughes16}. 

The study of extended bodies in general relativity was pioneered by Mathisson,  Papapetrou and Dixon et al. \cite{mpdm,mpdp,dixon64,dixon70a,dixon70b,dixon73,dixon74,dixon79}, the equations of motion now are known well as Mathisson-Papapetrou-Dixon (MPD) equations. The MPD equations describe the motion of an extended body with spin and arbitrary mass-multipoles in curved space-time. However, the MPD equations are not in closed forms. One needs a normalization condition and a spin supplementary condition to get a velocity-momentum relation. An early work was done by Ehlers and Rudolph \cite{ehlers77}. Recently, Xie and Kopeikin gave a canonical relation in the scalar-tensor theory \cite{kopeikin14}. Obviously, the trajectory of extended body deviates from the geodesic motion of the test particle because of the highly nonlinear Papapetrou force. A lot of literature have studied on the dynamics of spinning particles orbiting black holes, for example, revealing the orbital properties and chaos of the particles with unrealistically large spins ( see \cite{suzuki97,semerak99,suzuki00,hartl03,hartl04,bini04,semerak07,han08} and references inside).  

Considering that the quadrupole is the main part of multipoles in astrophysics and celestial mechanics, recently, several authors began to use MPD equations to research the dynamics of extended bodies up to quadrupoles, such as extended bodies with spin-induced quadrupoles in Schwarzschild and Kerr spacetimes \cite{bini13, bini15, vines16}, the bodies with spin- and tidal- induced quadrupoles around Kerr black holes \cite {steinhoff12}, and the ones with generic quadrupoles in Kerr spacetime\cite{bini14}. All of these literature are based on the MPD equations. In \cite{steinhoff12}, the authors derived an effective potential and bind energy for the body moving on the equatorial plane of a Kerr black hole, and they also compared their results with post-newtonian Hamiltonian in extreme mass ratio situations. Bini and Geralico studied the equatorial motion of extended bodies with a general quadrupole tensor \cite{bini14}. Bini et al. also studied the role of spin-induced quadrupole in the equatorial motion around Schwarzschild and Kerr blackholes \cite{bini13, bini15}, they studied the radial effective potential and analytically determined the innermost stable circular orbit (ISCO) shift. Most recently, Vines et al. derived a canonical Hamiltonian for an extended spinning test body in a curved background spacetime.

Most of the above works tried to give some analytical results based on small spin or post-Newtonian approximation, and the analytical and numerical calculations were limited on the equatorial motion (2D orbit). In the present paper, for the extended bodies orbiting the Kerr black holes, we give a closed form of the momentum-velocity relation, and then some numerical values of orbital frequencies of the simplest but the most important case: circular orbits. The magnitudes of frequency differences decide if we need consider the spin-induced quadrupoles in the GW templates for the low frequency GW detectors or not. We constrain the orbits on the equatorial plane to study the relation between rotating velocity and orbital radius of the small body. We then scan the parameter space to demonstrate the generic orbits of the extended bodies with artificially large spins and quadrupoles by numerical integration of the MPD equations. Specially, we study the influence of quadrupoles on the appearance of chaotic orbits. For the first step of a series of  researches, in this paper, we assume that the quadrupoles are only induced by the spins.  From the numerical results, we conclude if or not the chaos can happen for physical spins and quadrupoles. 

Through this paper, we use units where $G = c = 1$ and sign conventions ($-, +, +, +$). The time and space scale is measured by the mass of black hole $M$, energy and the linear momentum of particle are measured by it¡¯s mass $m$, the angular momentum and spin by $m M$. 

\section{Mathisson-Papapetrou-Dixon equations}
Mathisson and Papapetrou gave the pole-dipole approximation description for spinning particles \cite{mpdm, mpdp}, and Dixon extended it to arbitrary multipole moments \cite{dixon79}.  Now these equations are called as MPD equations  for the motion of extended bodies in general relativity. The equations are written as:
\begin{align}
\dot{p}^\mu &= -\frac{1}{2} R^\mu_{\nu\alpha\beta} \upsilon^\nu S^{\alpha\beta} - F^\mu,  \label{mpd1} \\
\dot{S}^{\mu\nu} &= 2p^{[\mu}\upsilon^{\nu]} + F^{\mu\nu} , \label{mpd2}
\end{align}
where dot means covariant differential, $p^\mu=m u^\mu$ is the total momentum, $S^{\alpha\beta}$ the anti-symmetrical spin tensor, $m$ the dynamical mass, $\upsilon^\mu = d x^\mu /d\tau$ the kinematical four-velocity, $\tau$ an arbitrary affine parameter, and when taking up to the quadrupole, we have
\begin{align}
F^\mu &\equiv \frac{1}{6} J^{\alpha\beta\gamma\sigma}\nabla^\mu R_{\alpha\beta\gamma\sigma}, \\
F^{\mu\nu} & \equiv \frac{4}{3}J^{\alpha\beta\gamma[\mu} R^{\nu]}_{\gamma\alpha\beta} ,
\end{align}  
where $J^{\alpha\beta\gamma\sigma}$ is the quadrupole tensor.

The above equations are not in closed forms. The center of mass of the extended body and the affine parameter $\tau$ should be decided by two supplementary conditions. To determine the center of mass, the popularly adopted supplementary condition is $u_\nu S^{\mu\nu} = 0$ \cite{tulczyjew59, dixon64} which is also used in this paper (sometimes $\upsilon_\nu S^{\mu\nu} = 0$ is adopted in few literature, see Ref. \cite{semerak07} for a comparative discussion). The orthogonal condition $u^\mu \upsilon_\mu = -1$ is employed to determine the parameter $\tau$. This condition was given in \cite{ehlers77} and was used by most researchers such as \cite{bini13,bini15,vines16,steinhoff12,bini14,kopeikin14}. Generally, the dynamical mass $m$ is not a constant, and can be determined by $p^\mu p_\mu = -m^2$. Based on the former definitions, it yields $u^\mu u_\mu = -1$. Following the similar procedure in \cite{semerak99} for the spinning test particles, from Eq. (2), we have 
\begin{align}
p_\nu \dot{S}^{\mu\nu} = m^2 \upsilon^\mu - m p^\mu + F^{\mu\nu} p_\nu \label{pdots}, 
\end{align}
and 
\begin{align}
\upsilon^\nu =  m^{-2} ( p_\mu \dot{S}^{\mu\nu} + m p^\nu - F^{\mu\nu} p_\mu ).
\end{align}
Taking Eq. (6) to (1), and times $S^{\sigma\mu}$ we have
\begin{align}
2 m^2 \dot{p}_\mu S^{\sigma\mu} = -m R_{\mu\nu\alpha\beta} p^\nu S^{\sigma\mu} S^{\alpha\beta} - R_{\mu\nu\alpha\beta} {p}_\delta \dot{S}^{\nu\delta} S^{\sigma\mu} S^{\alpha\beta} + R_{\mu\nu\alpha\beta} F^{\nu\delta} p_\delta S^{\sigma\mu} S ^{\alpha\beta}- 2 m^2 F_\mu S^{\sigma \mu}. 
\end{align}
Using 
\begin{align}
R_{\mu\nu\alpha\beta} S^{\nu\delta} S^{\sigma\mu} &= -\frac{1}{2} R_{\mu\nu\alpha\beta} S^{\sigma\delta} S^{\mu\nu}, \\
\dot{p}_\delta S^{\nu\delta} &= - p_\delta \dot{S}^{\nu\delta} ,
\end{align}
together with Eq. (5), finally we get the relation between the four-velocity and the linear momentum:
\begin{align}
m^2 \upsilon^\sigma = m p^\sigma - F^{\sigma \nu} p_\nu+ \frac{2m R_{\mu\nu\alpha\beta} p^\nu S^{\sigma\mu} S^{\alpha\beta} - 2 R_{\mu\nu\alpha\beta} F^{\nu\delta} p_\delta S^{\sigma\mu} S ^{\alpha\beta}+ 4 m^2 F_\mu S^{\sigma \mu}}{4m^2 + R_{\mu\nu\alpha\beta} S^{\mu\nu} S^{\alpha\beta}}. \label{vprelation}
\end{align}
The above equation is equivalent with the relation (2.17) given by Ehlers and Rudolph \cite{ehlers77}.  However, our relation may be more convenient for practice because the geometric quantities involved are only metric and Riemann curvature tensor. One can choose $\upsilon^\mu \upsilon_\mu =-1$ to make the $\tau$ as the proper time, and obtain the kinematical mass $\bar{m} \equiv p^\mu \upsilon_\mu$ (usually $\bar{m} \neq m$, see \cite{steinhoff12} for details). In the present paper, we adopt the orthogonal condition $u^\mu \upsilon_\mu = -1$ (following Ehlers and Rudolph), then $m = \bar{m}$, from which calculations are simplified and get the explicit relation (\ref{vprelation}).  

\section{circular orbits on equatorial plane}
The circular orbits are very important when we discuss the motion around black holes. For example, by observing orbital frequency from GW signals or pulsar timing, one can determine the spin and quadrupole values of the compact objects, then constrain the equations of states of them. For the test particles, there is a simple but exact expression for determining  circular orbits. 
For pole-dipole approximation, the case becomes complicated, the expression is quite long but still acceptable \cite{suzuki00, han10}.  However, when considering the extended bodies,
 we numerically solve out the circular condition. Numerical solving is easier and accurate, though in principle, one can also find out a very complicated analytical solution.  
\subsection{Initial data determination}
Based on the analysis in \cite{steinhoff12}, the circular orbits request $\upsilon^r = u^r = 0, ~\upsilon^\theta = u^\theta = 0$  and two other conditions with given radius $r$:
\begin{align}
& u^t u_t + u^\phi u_\phi= -1; ~ \dot{p}^r (u^t, u^\phi) = 0\,,  \label{circular}
\end{align}
where dot means $d/d\tau$. From these constraints one in principle can give analytical expressions of the initial conditions for $u^t, ~u^\phi$ and orbital frequency $\Omega_\phi$. However, the expressions are too long for conveniently using. It is more convenient to search the roots of Eqs. (\ref{circular}) numerically. For example, in this paper, we find out the solution of  Eqs. (\ref{circular}) with Monte-Carlo root finder routine by setting an error less than $10^{-15}$.  

It is useful to introduce the spin vector $S^\mu$,
\begin{align}
S_\mu = -\frac{1}{2} \epsilon_{\mu\nu\alpha\beta} u^\nu S^{\alpha\beta} \,.
\end{align}
The only nonzero component of spin for circular orbits is $S^\theta$. The inverse transformation is
\begin{align}
S^{\mu\nu} =  \epsilon^{\mu\nu\alpha\beta} u_\alpha S_{\beta}, 
\end{align}
where $\epsilon^{\mu\nu\alpha\beta} = \varepsilon^{\mu\nu\alpha\beta}/\sqrt{-g}$ is a tensor and $\varepsilon^{\mu\nu\alpha\beta}$ the Levi-Civita alternating symbol ($\varepsilon_{0123} \equiv 1,~\varepsilon^{0123} \equiv -1$).  

A spin magnitude $S$ can be introduced as
\begin{align}
S^2 = S^\mu S_\mu = \frac{1}{2} S^{\mu\nu} S_{\mu\nu} . \label{sconstant}
\end{align}
we then have $S^\theta = - S/\sqrt{g_{\theta\theta}}$. As a result, we have only 4 nonzero components of the spin tensor for circular-equatorial orbits, i.e.
\begin{align}
S^{tr} = - S^{rt} = -S u_\phi /r, ~ S^{r\phi} = - S^{\phi r} = -S u_t /r \,. 
\end{align}
Taking these quantities into the nonlinear Eqs. (\ref{circular}) of $u^t, ~u^\phi$, then using the Monte-Carlo root-finder method, we can solve out $u^t, ~u^\phi$ with errors less than $10^{-15}$ and then $\upsilon^t, ~\upsilon^\phi$ for the circular orbits. Numerical simulation shows that the initial data we got keep the circular orbits with machine precision.

\subsection{Orbital frequency}
The orbital frequency is defined as $\Omega_\phi \equiv d\phi/d t =v^\phi/v^t$. The main frequency of GWs radiated from the circular motion, is the double of orbital one. 
In \cite{han10}, we have shown that for spin $\sim 10^{-5}$, the frequency difference due to spin still need to be considered in GW detections.  Now let's see if the effects of quadrupoles are considerable or not. For the first step, we assume that the quadrupoles are induced by spin only, from \cite{bini15} we have
\begin{align}
J^{\alpha\beta\gamma\delta} = 4 u^{[\alpha} \chi(u)^{\beta][\gamma}u^{\delta]} \,,
\end{align}
with $\chi(u) = \frac{3}{4} \frac{C_Q}{m} [S^{\alpha\gamma} S_\gamma^{~\beta}]^{\rm STF}$ where $C_Q$ is a ¡°polarizability¡± constant. ``STF" means symmetrical-trace free part of a tensor, i.e., 
\begin{align}
[S^{\alpha\gamma} S_\gamma^{~\beta}]^{\rm STF} = S^\alpha S^\beta - \frac{1}{3} S^2 P(u)^{\alpha\beta} \,,
\end{align} 
where $P(u)^{\alpha\beta} \equiv g^{\alpha\beta} + u^\alpha u^\beta$ is the projection operator.  The values of $C_Q$ associated with compact objects are given, e.g., in Ref. \cite{hergt14}. The normalization is such that $C_Q = 1$ in the case of a black hole \cite{thorne80}, whereas for neutron stars $C_Q$ depends on the equation of state and varies roughly between 4 and 8 \cite{poisson99}. 

\begin{figure}
\begin{center}
\includegraphics[height=2.0in]{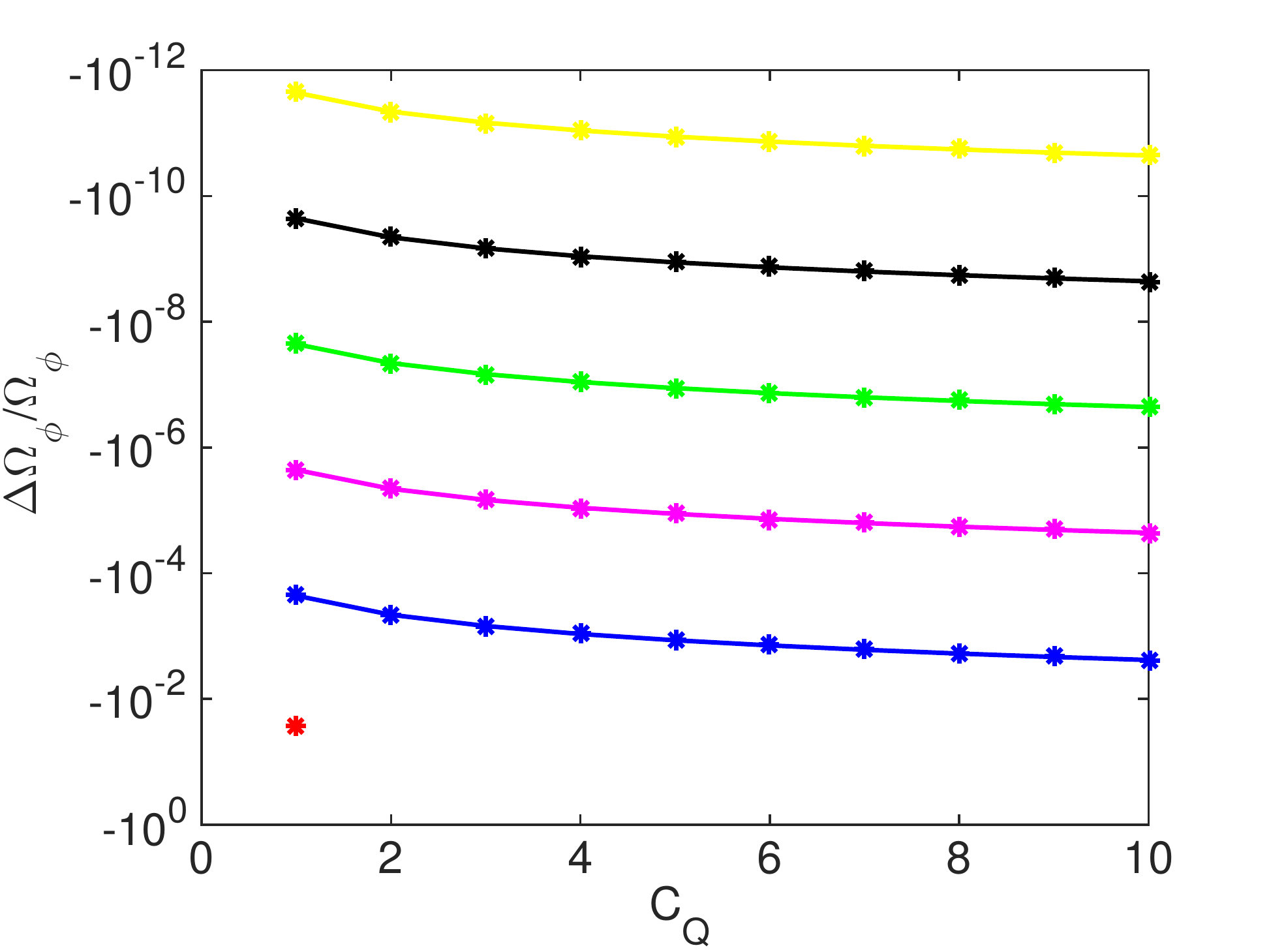}
\includegraphics[height=2.0in]{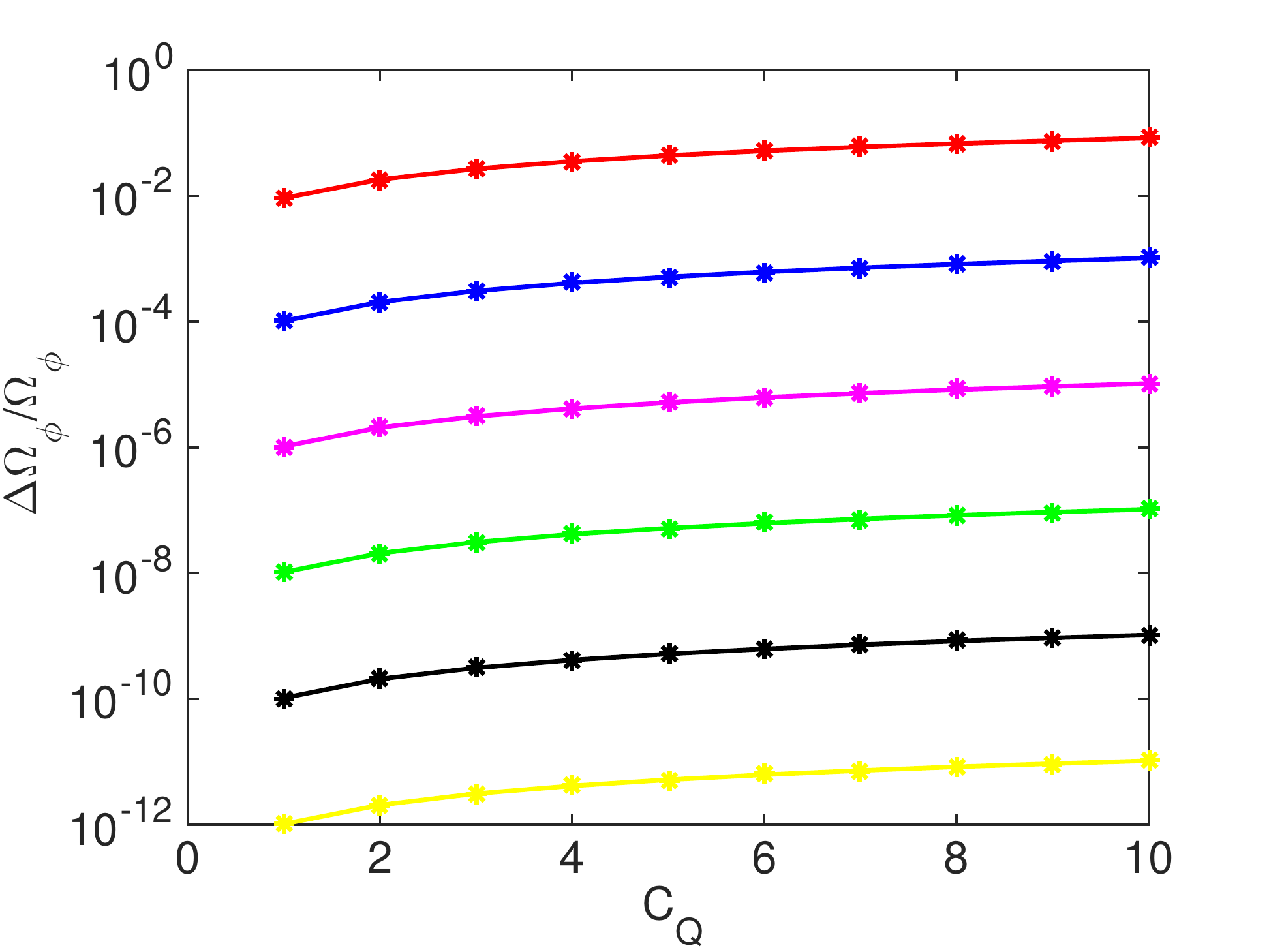}
\includegraphics[height=2.0in]{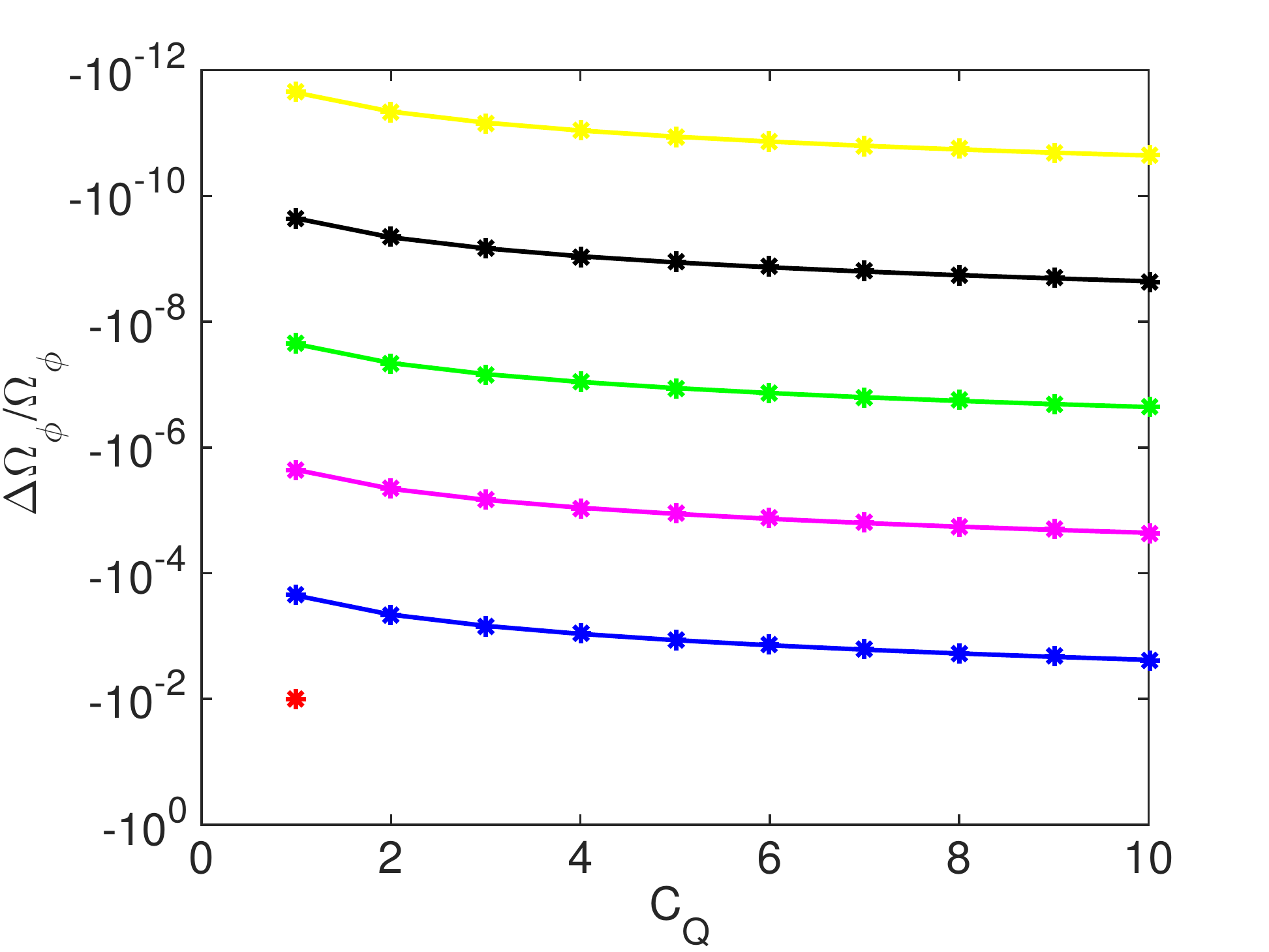}
\includegraphics[height=2.0in]{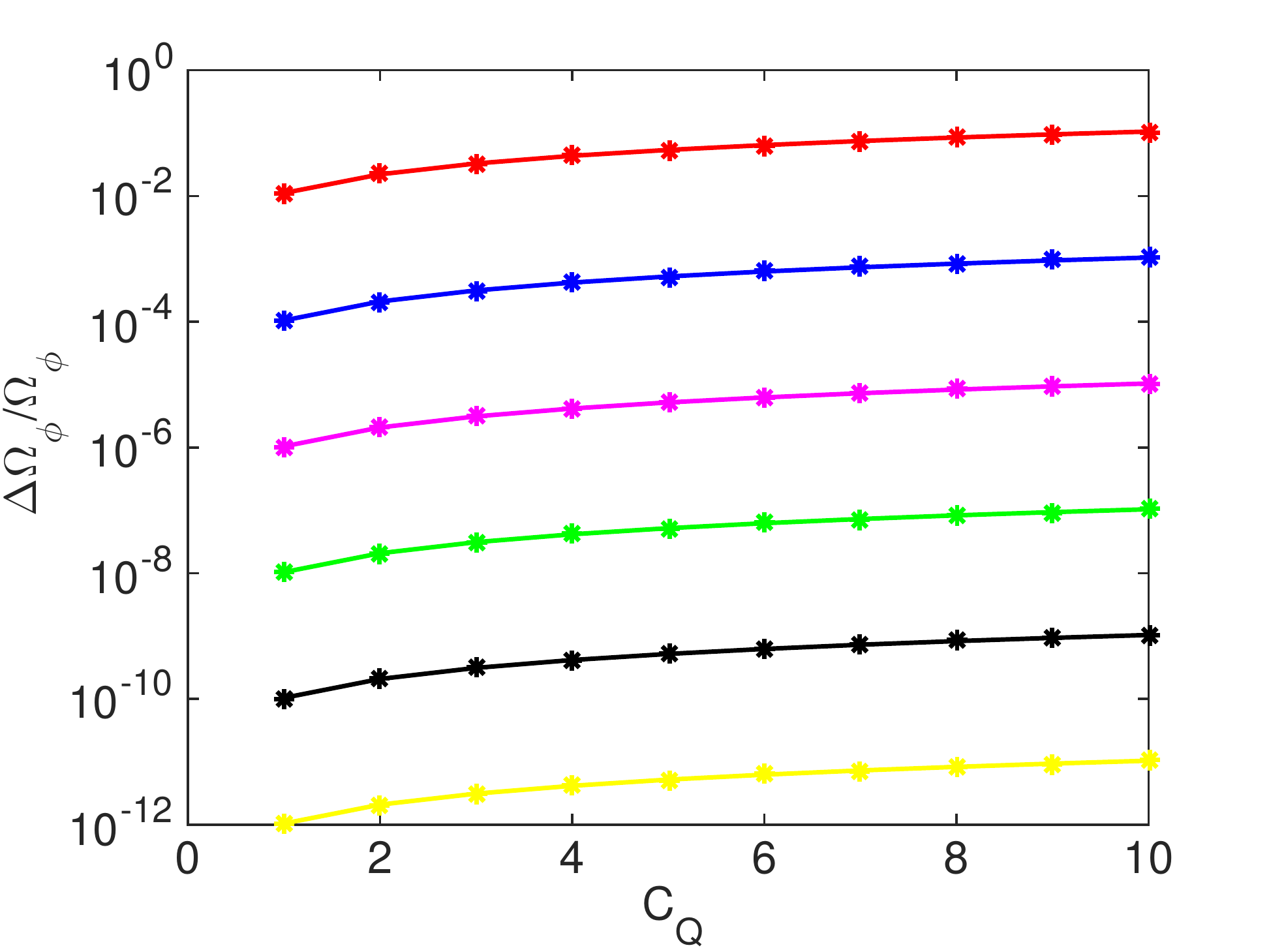}
\caption{The orbital frequency shifts of extended bodies circularly orbiting an extreme Kerr black hole ($a=1$). The above panels show the results of aligned spin cases ($S>0$), and the bottom panels the ones of anti-align spins ($S<0$). The orbital radius in the left (right) panels is $r_{\rm circ} =2$} ($6$). All the lines with red, blue, pink, green, black and yellow represents $S = 1$, $0.1$, $10^{-2}$, $10^{-3}$, $10^{-4}$ and $10^{-5}$ respectively.  \label{fig1}
\end{center}
\end{figure}

In Fig. \ref{fig1}, we draw the values of $\Omega_\phi (C_Q \neq 0) -\Omega_\phi (C_Q = 0)$ as a function of $C_Q$ with the different $S$ and $r_{\rm circ}$ for the extreme Kerr black hole.  
We can do a simple estimation for the dephase due to this frequency shift. In a highly relativistic regime, the large mass-ratio binaries inspiral about $M/m$ cycles, with the typical orbital frequency of $O(10^{-2}\frac{1}{M})$, then the dephase  is estimated as 
\begin{align}
\Delta \phi \sim \Delta \Omega_\phi \frac{M}{m} \frac{2\pi}{\Omega_\phi} \,.
\end{align}

In our normalized units, the spin parameter $S$ is measured in terms of $mM$, not $m^2$. The system we consider in this paper is a compact extended body of mass $m$ orbits a large body of mass $M$. The mass-ratio $m/M$ must $\ll 1$ to make sure we can omit the gravitational self-force of the small body and then the MPD equations maintain valid. If the small one is a black hole,  we know that a maximally spinning black hole of mass $m$ has spin angular momentum $m^2$. Therefore for a small black hole $m$ orbiting a large black hole of mass $M$, the spin parameter $S \sim m^2/(mM) = m/M \ll 1$. For neutron stars and especially white dwarfs, the situation becomes more complicated. Based on the detailed analysis in Ref. \cite{hartl03}, in all cases, the realistic $S$ should $\ll 1$. 

In Fig. \ref{fig1}, we can see clearly that for the physical spin ($S < 10^{-2}$) of large mass-ratio systems, the frequency shifts are too small to be considered. As a reference, the values of orbital frequencies are listed in Table \ref{tab1}.

\begin{table}
\caption{The orbital frequency  $\Omega_\phi$ of extended bodies while $r_{\rm circ} =6$ and
 $a=1$. Attend that the numbers in the table need multiply $10^{-2}.$}\label{tab1}
\begin{tabular}{c|c c c c c c}
\hline \hline
$S$&$-10^{-3}$&$-10^{-4}$&$-10^{-5}$&$10^{-5}$&$10^{-4}$&$10^{-3}$\\
\hline $C_Q = 0$&$6.371029346056$&$6.370705074859$&$6.370672652106$&$6.370665447158$&$6.370633025376$&$6.370308851228$\\
$C_Q = 1$&$6.371029412349$&$6.370705075522$&$6.370672652113$&$6.370665447165$&$6.370633026039$&$6.370308917501$\\
$C_Q = 4$&$6.371029611226$&$6.370705077510$&$6.370672652133$&$6.370665447184$&$6.370633028028$&$6.370309116320$\\
$C_Q = 6$&$6.371029743811$&$6.370705078836$&$6.370672652146$&$6.370665447198$&$6.370633029353$&$6.370309248866$\\
$C_Q = 8$&$6.371029876396$&$6.370705080161$&$6.370672652159$&$6.370665447211$&$6.370633030679$&$6.370309381412$\\
 \hline \hline
\end{tabular}
\end{table}

\begin{figure}
\begin{center}
\includegraphics[height=2.0in]{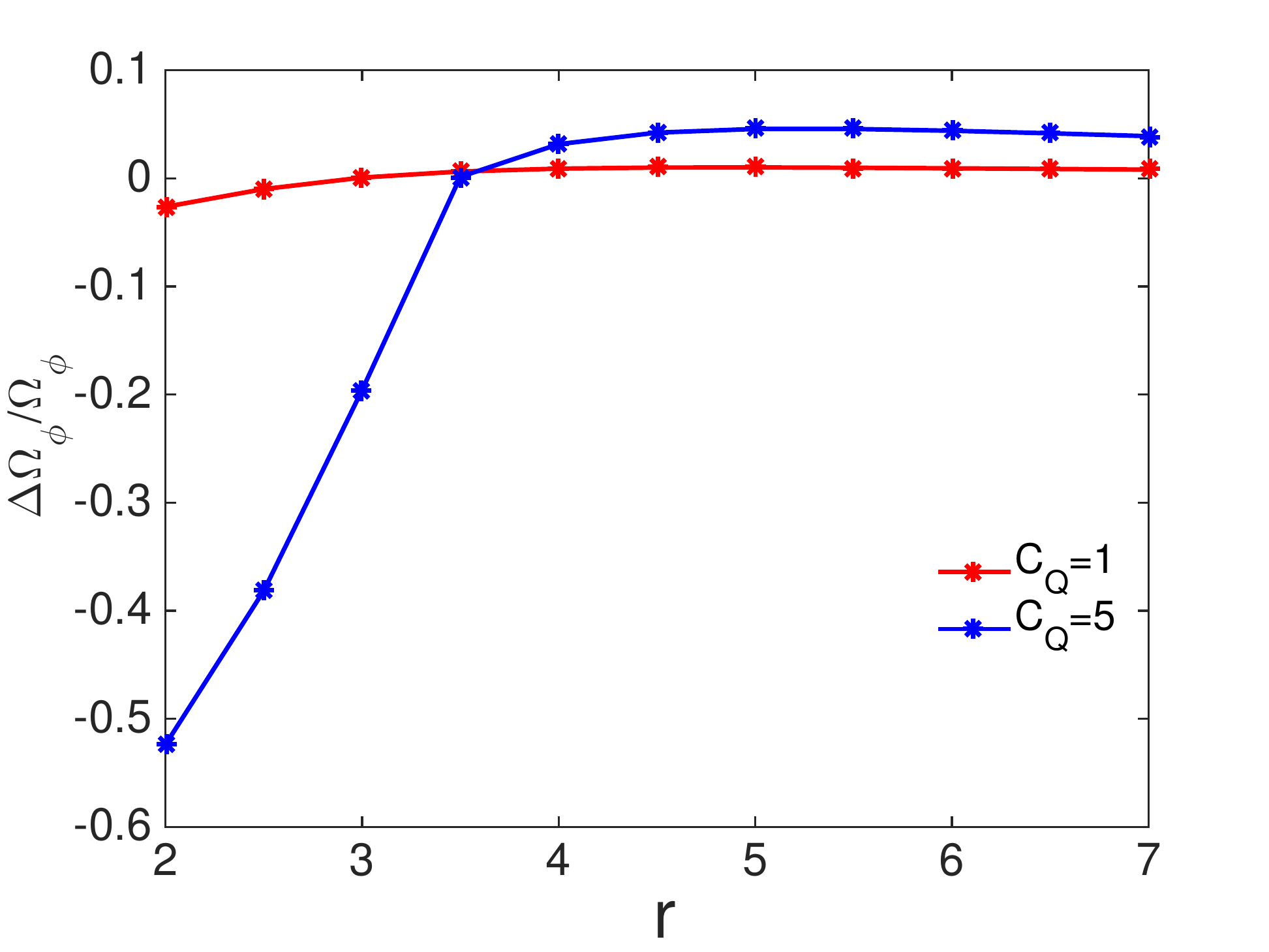}
\includegraphics[height=2.0in]{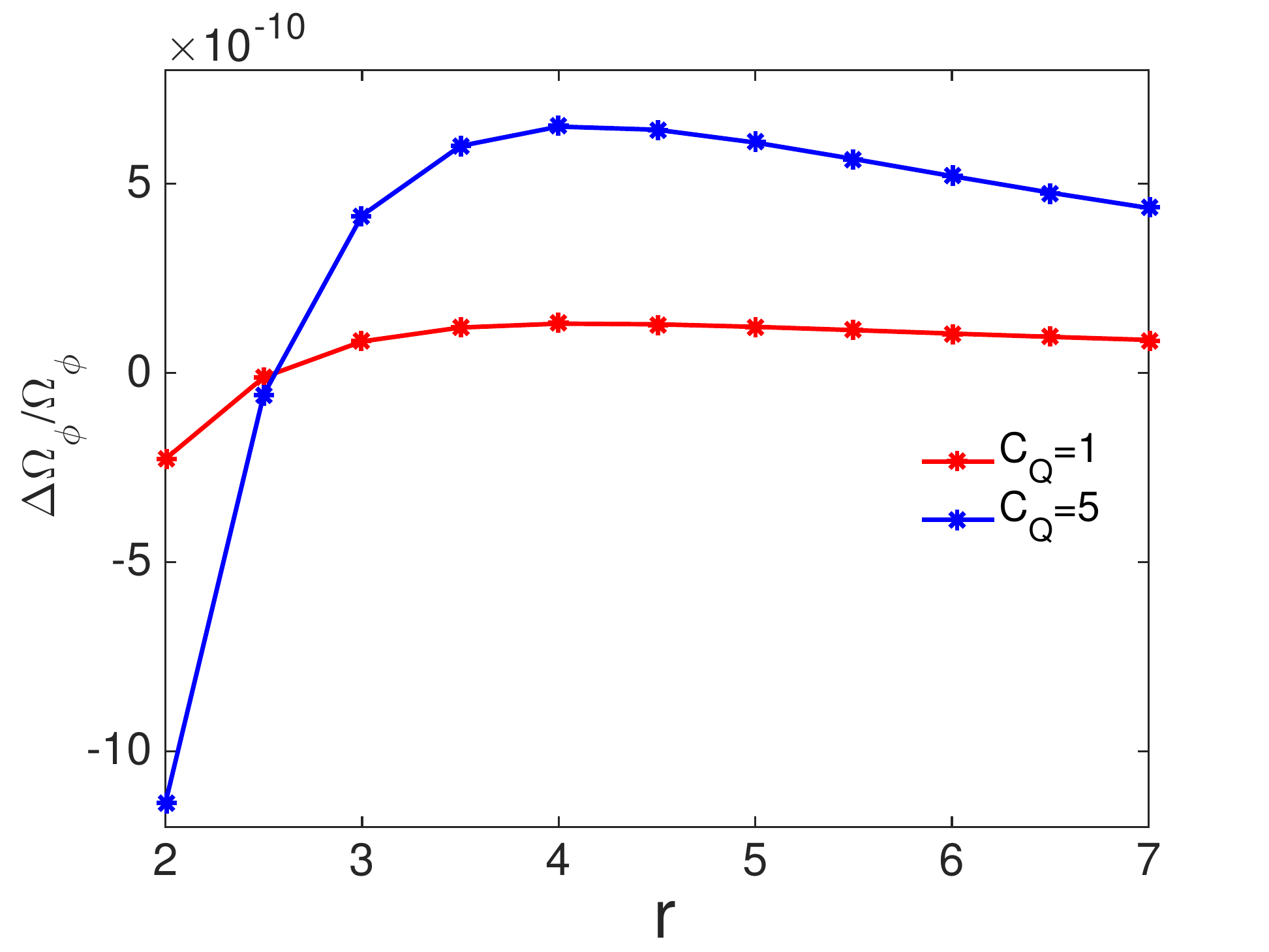}
\caption{The orbital frequency shifts of extended bodies circularly orbiting an extreme Kerr black hole ($a=1$). The left and right panels show the results for aligned spin value $S=1$ and $S=10^{-4}$ respectively. } \label{fig2}
\end{center}
\end{figure}

Obviously, the differences of frequency depend on the values of spin and $C_Q$. $|\Delta \Omega_\phi/\Omega_\phi|$ will be increased with the increase of $S$ and $C_Q$. We find that it is almost a linear relation between $|\Delta \Omega_\phi/\Omega_\phi|$ and $C_Q$. This linear relation does not be revealed intuitively in Fig. \ref{fig1}, because we use a log plot. However, the frequency shift due to a fixed 
$C_Q$ is a function of orbital radius. From Fig. \ref{fig2}, when $r_{\rm circ}$ is large enough, the influence of quadrupole on orbital frequency become smaller while the extended body is farther from the black hole. There is a $r_{\rm t}$ where the $\Delta\Omega_\phi$ arrives at its maximum. When $r_{\rm circ}$ is smaller than $r_{\rm t}$, $\Delta \Omega_\phi$ decreases to zero at first and then becomes negative. After the zero point, the absolute value of $\Delta \Omega_\phi$ is larger and larger while the body is closer and closer to the black hole. The details can be found in Fig. \ref{fig2}.

\section{Generic orbits}
\begin{figure}
\begin{center}
\includegraphics[height=2.0in]{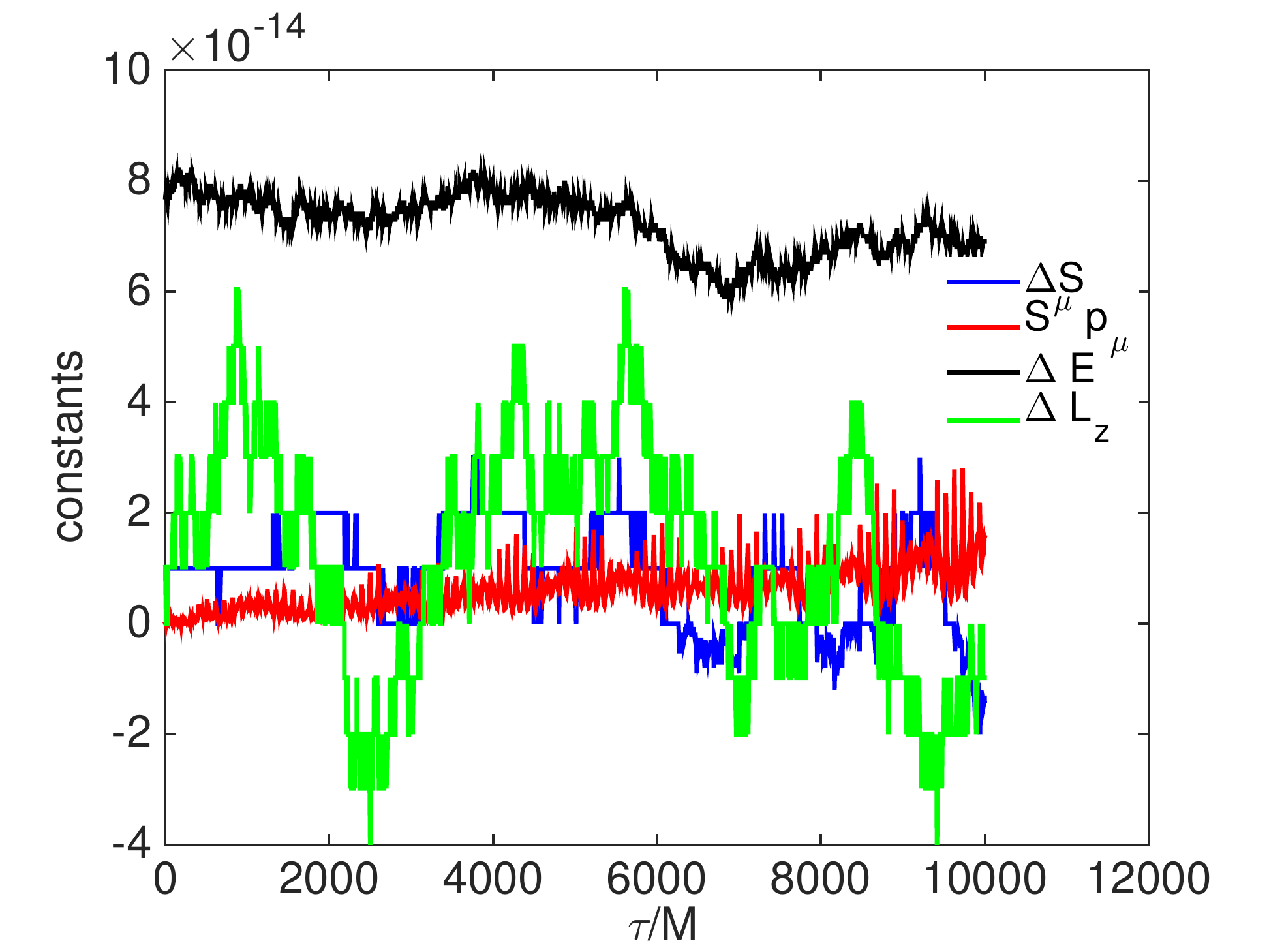}
\caption{Numerical performance of the kinematical constants of an extended body during orbital evolution. The parameters used are: $a=1, E=0.8837, L_z=2.8, S=1, C_Q=2$. The body is put at $r=6, \theta =\pi/2, \phi =0$ and $p^\theta =0$ at beginning. Here $\Delta S = S(\tau)-S_0$ and so on. } \label{fig3}
\end{center}
\end{figure}

For generic orbits in three dimensions, there are total 12 orbital parameters need to be given as the initial data at the beginning: $x^\mu, ~u^\mu, ~S^\mu$. Considering existing five constants or conservations, we have 7 orbital parameters left to be set. Usually, we choose the following initial conditions of the particle: $t_0, r_0, \theta_0, \phi_0$, $u^r_0$ or $u^\theta_0$, and $S^r_0, S^\theta_0$. The other five initial parameters will be solved out from five conservation equations: 
\begin{align}
&u^\mu u_\mu =-1 , \\
&p^\mu S_\mu =0 , \\
&S^\mu S_\mu =S^2 , \\
&E = -p_t + \frac{1}{2} g_{t \mu , \nu} S^{\mu\nu} , \label{energy} \\
&L_z  = p_\phi - \frac{1}{2} g_{\phi \mu , \nu} S^{\mu\nu}  \label{angular} ,
\end{align}
where $E, ~L_z$ are the energy and total angular momentum of the body. So, giving $E, ~L_z$ and spin magnitude, together with 7 initial orbital parameters, the 
motion of an extended body is fully determined. Of course, we need numerically solve the MPD equations together with the velocity-momentum relation (\ref{vprelation}). The numerical results can be validated by checking if  the conservations and constants are maintained during the orbital evolution or not. We find that with the condition $u^\mu \upsilon_\mu =-1$ and the velocity-momentum relation (\ref{vprelation}), the numerical evolution can keep all constants and constraints with very small errors (only machine random errors in our evolution).  Please see Fig. \ref{fig3} and Table \ref{tab2} for details. Therefore, we can conclude that our numerical codes are correct and accurate enough. 

\begin{table}
\caption{The values of constants after numerical evolutions to $\tau =10 ^4 $ M. All parameters are the same with the ones in Fig. \ref{fig3}. }\label{tab2}
\begin{tabular}{c c c c c}
\hline \hline
$\upsilon^\mu u_\mu $&$\Delta S$ & $S^\mu p_\mu$&$\Delta E$&$\Delta L_z$ \\
\hline
$-1$ &$-1.4\times10^{-14}$& $1.6\times10^{-14}$&$-6.9\times10^{-14}$&$-1.0\times10^{-14}$\\
 \hline \hline
\end{tabular}
\end{table}

\subsection{Equatorial orbits}
\begin{figure}
\begin{center}
\includegraphics[height=2.0in]{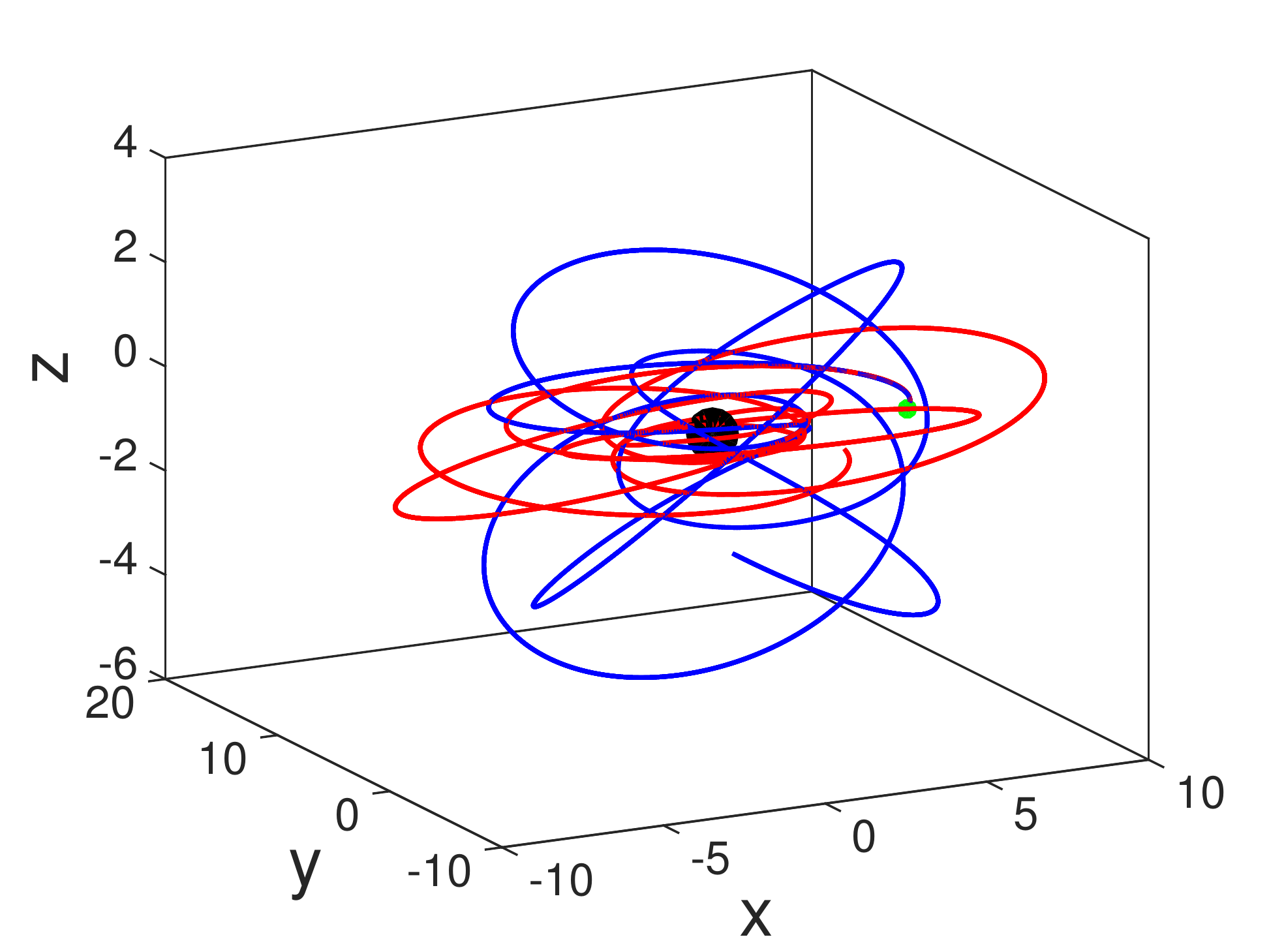}
\includegraphics[height=2.0in]{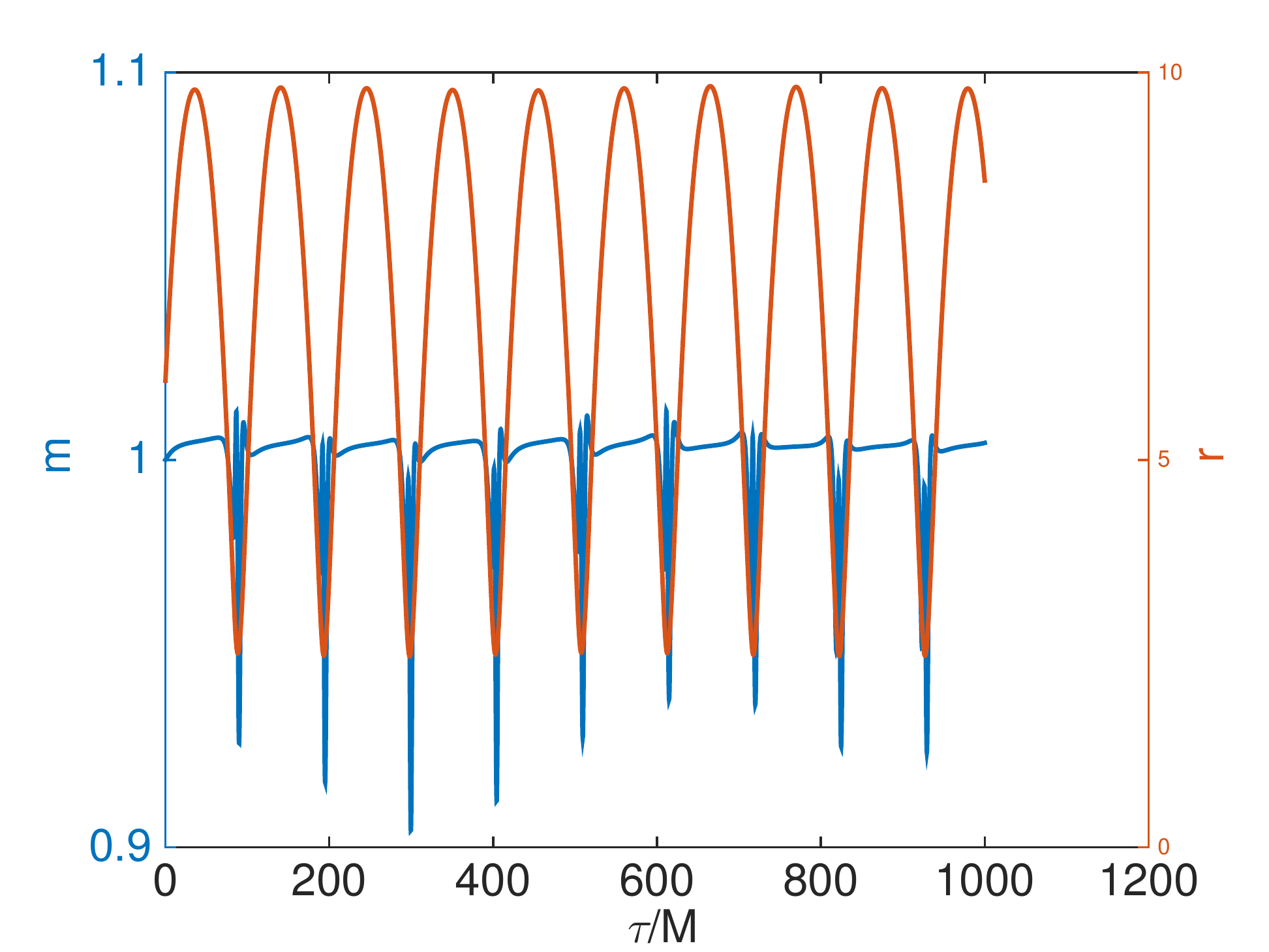}
\caption{Left panel: 3D orbits of two bodies with parameters $E=0.9237, L_z=2.8, S=1, C_Q = 2$ (red) and $C_Q = 0$ (blue). Both the two bodies are put at $r=6, \theta =\pi/2, \phi =0$ and $p^\theta =0$ at beginning (green point). Right panel: the dynamical mass of body and orbital radius.} \label{fig4}
\end{center}
\end{figure}

As we know, the dynamical mass $m \equiv \sqrt{-p^\mu p_\mu}$ is no longer a constant for an extended body. The variation of mass basically depends on the variation of orbital radius. Please see the right panel of Fig. \ref{fig4} for visualization. In the left one of this figure, we demonstrate the 3D trajectories of two particles with the same initial data and parameters but one with $C_Q = 2$ and the other $C_Q = 0$. We can see these two trajectories are totally separated after a short evolution. The mass of the extended body varies very sharply when it pass through the perihelion.  Considering the variation of mass mainly depends on the orbital radius, we constrain the orbits on the equatorial plane to do a detailed research. 

For the equatorial orbits, the only nonzero component of spin vector is $S^\theta = -S/\sqrt{g_{\theta\theta}}$, and the particle should be constrained on the equatorial plane with $\upsilon^\theta = p^\theta = 0$, but without $\dot{p}_r = 0$ at beginning.  In this case, the dynamical mass have a direct relation with the orbital radius. We find that for a certain equatorial-eccentric orbit, the mass increases when the orbital radius becomes larger. When the particle is at the perihelion (aphelion),  the value of $m$ is minimal (maximal). The variations of $m$ and $r$ share the same period.

\begin{figure}
\begin{center}
\includegraphics[height=2.0in]{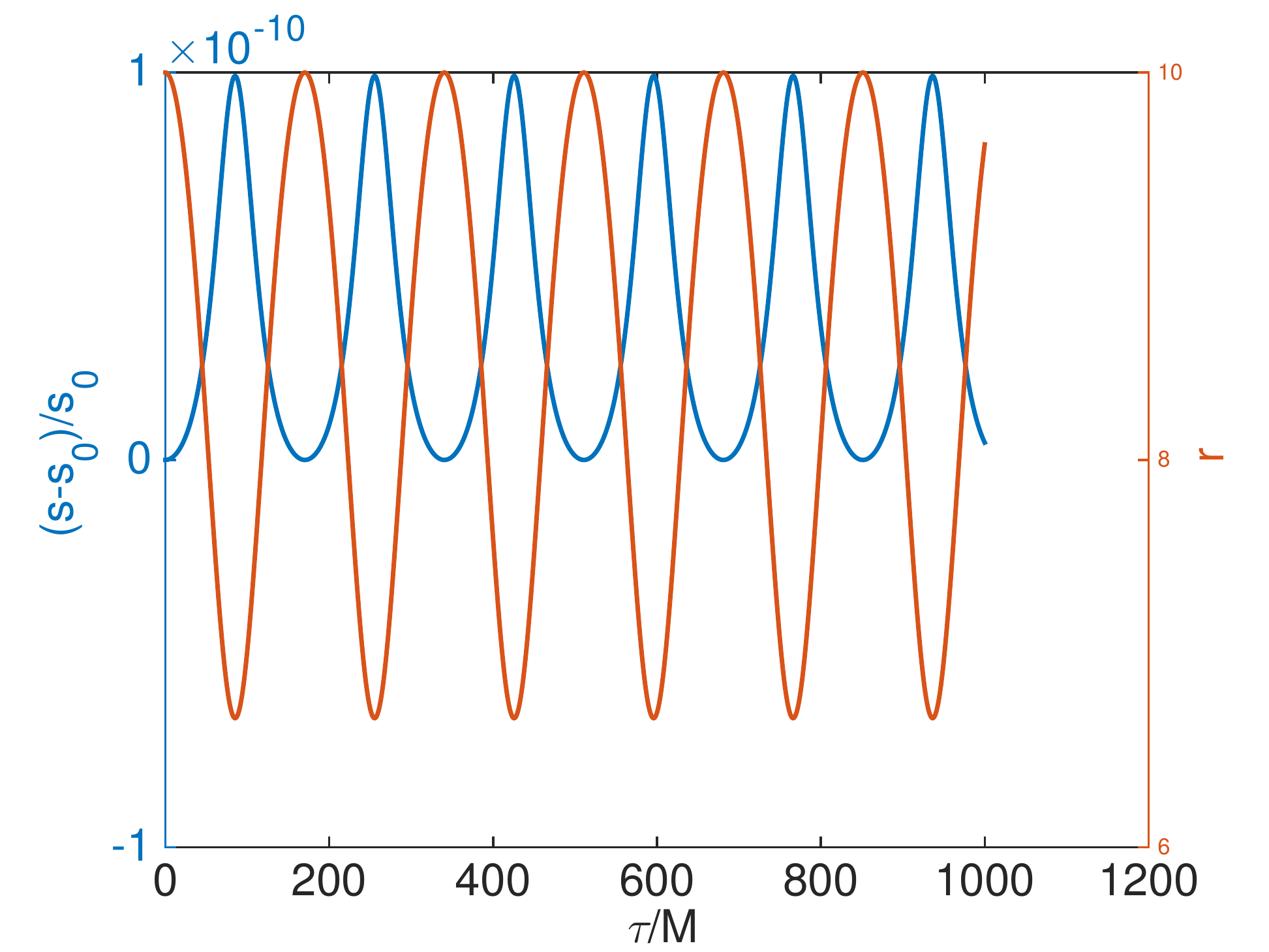}
\includegraphics[height=2.0in]{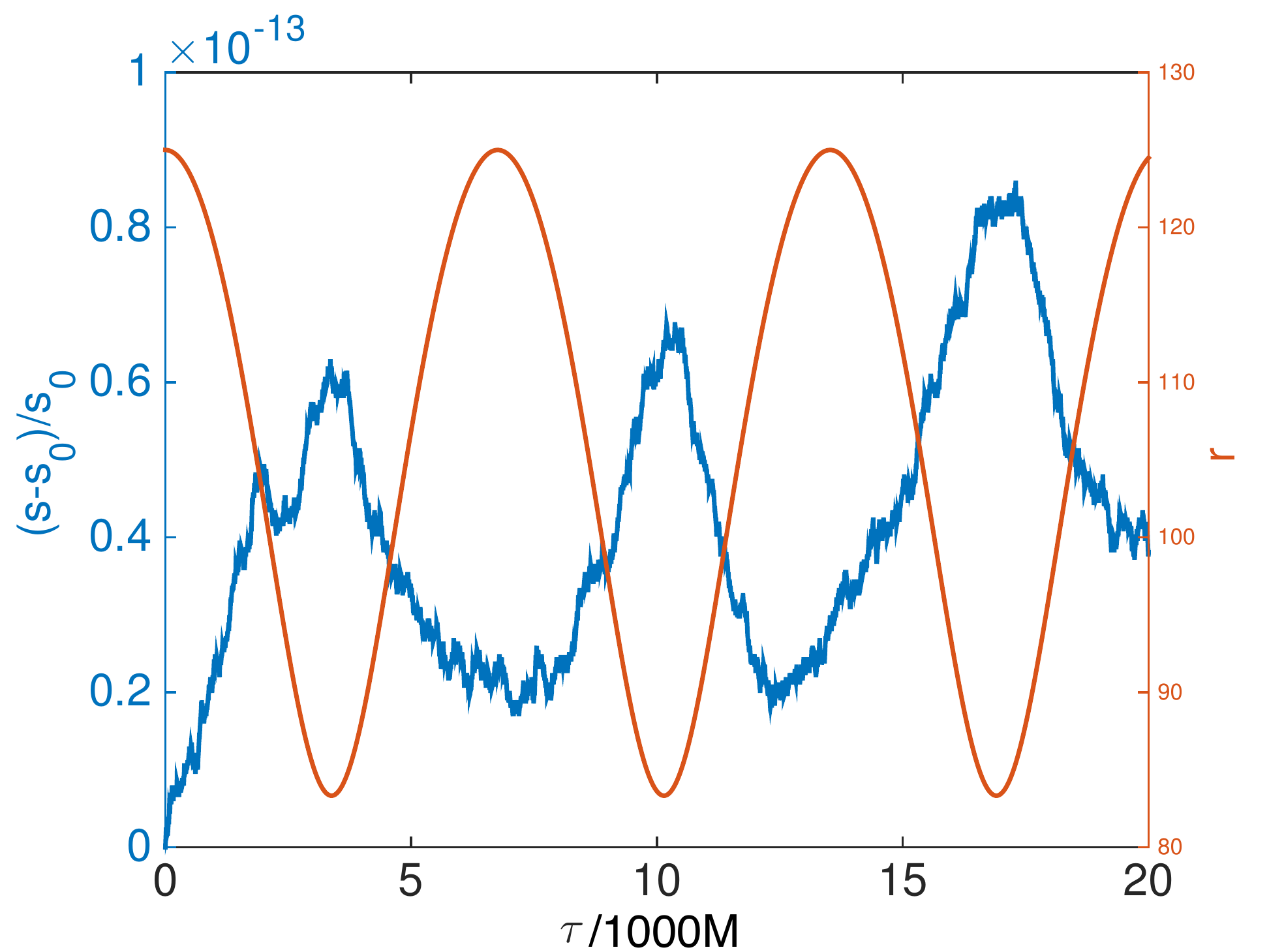}
\includegraphics[height=2.0in]{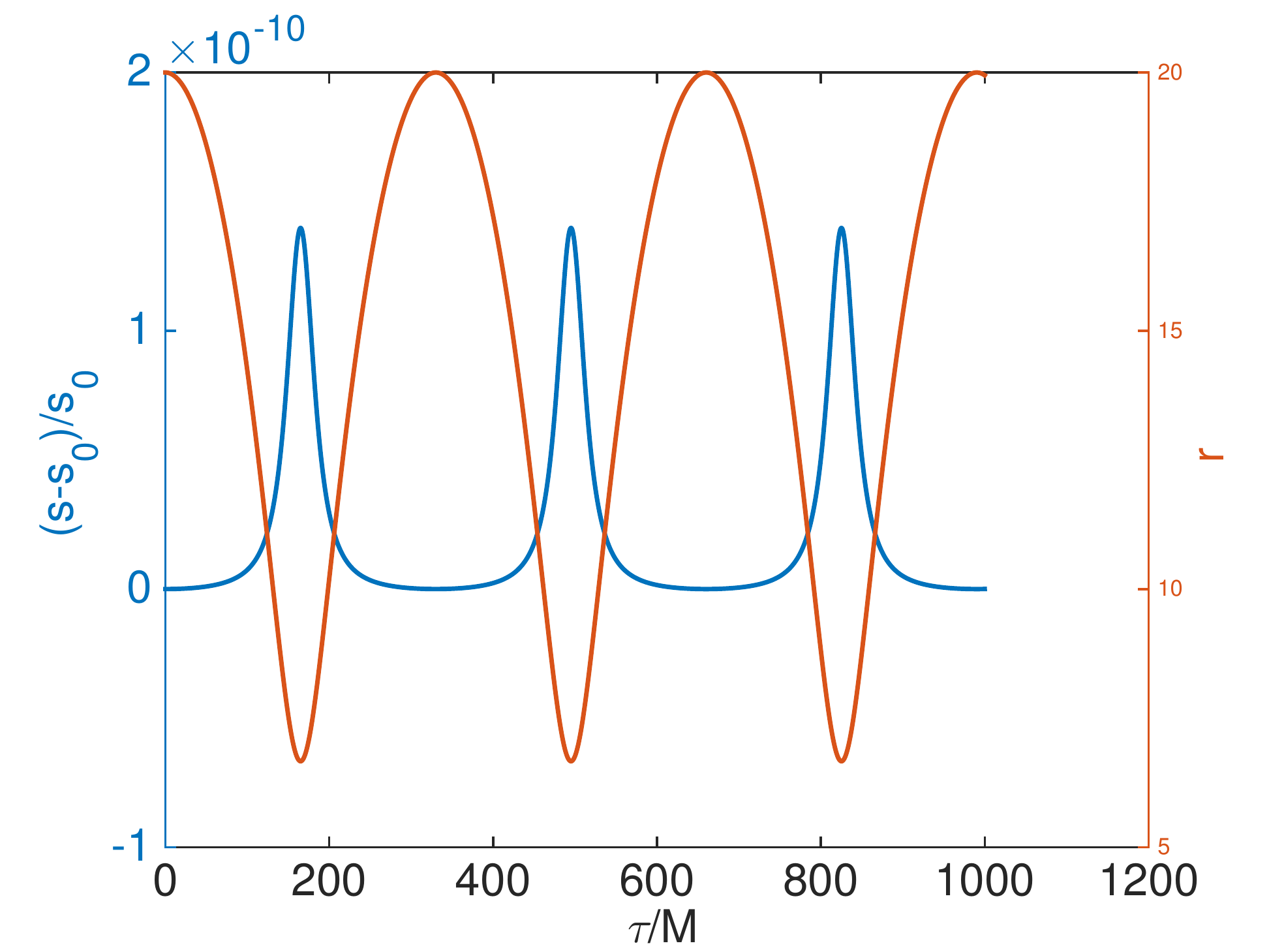}
\includegraphics[height=2.0in]{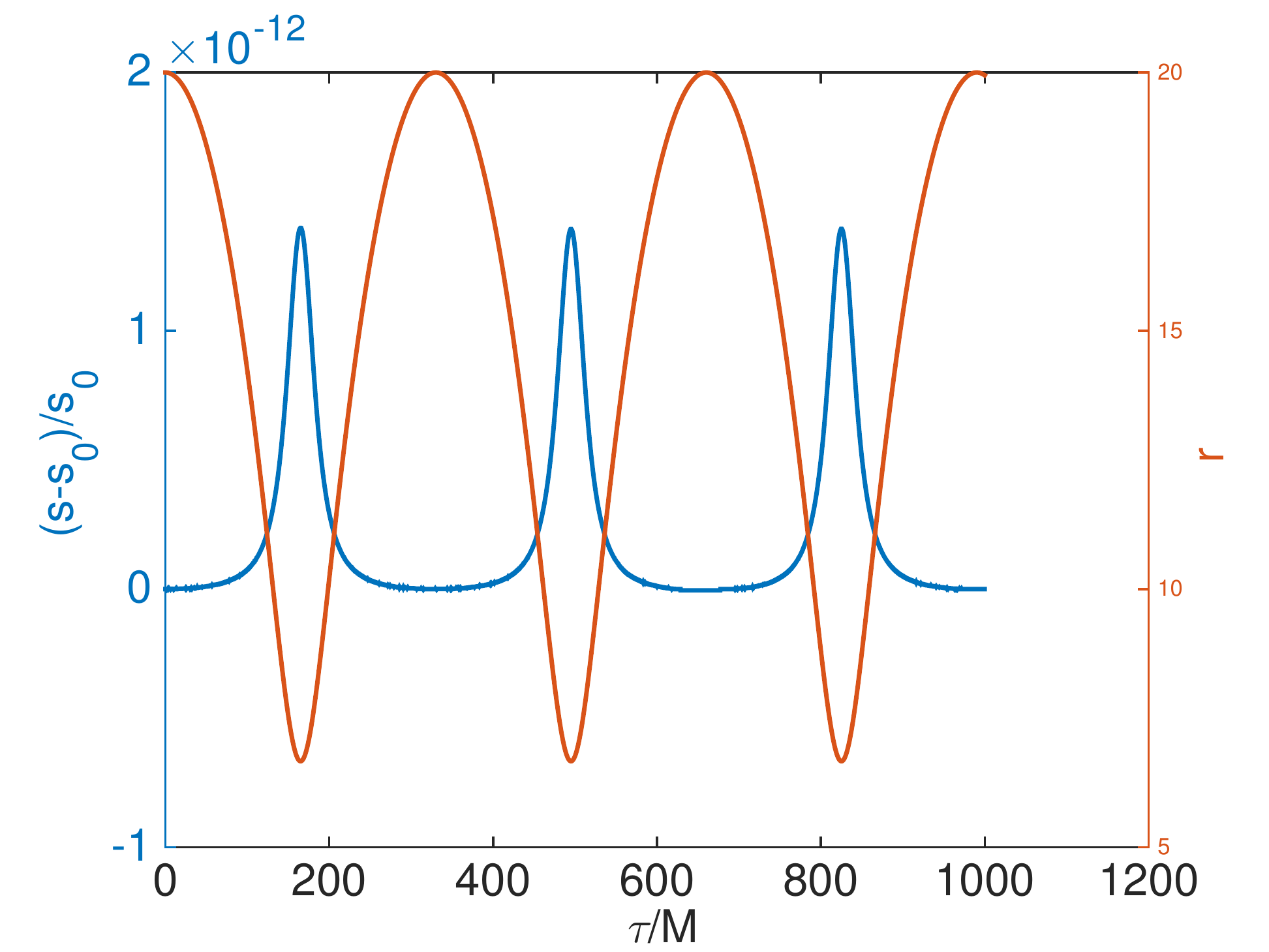}
\includegraphics[height=2.0in]{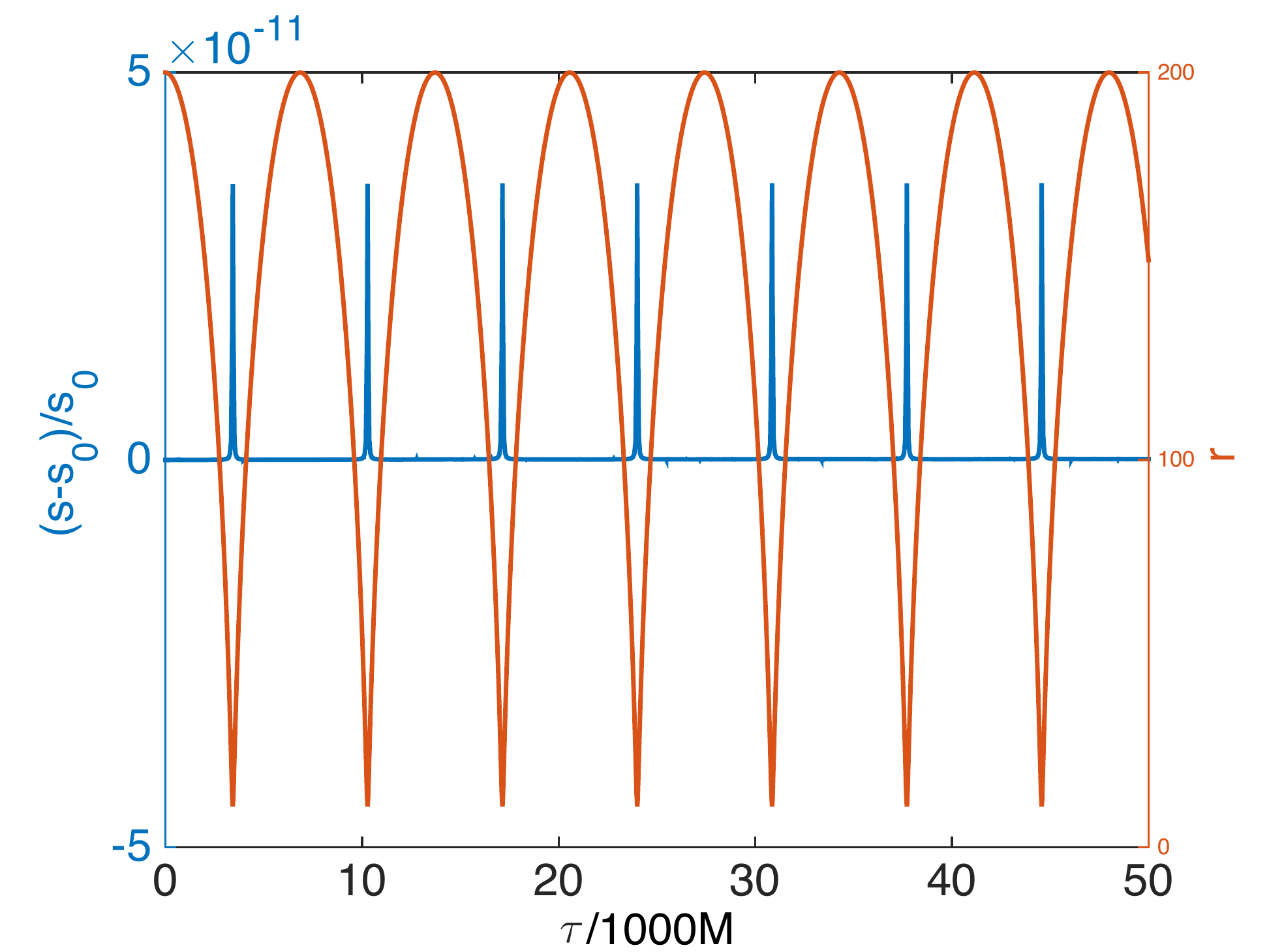}
\includegraphics[height=2.0in]{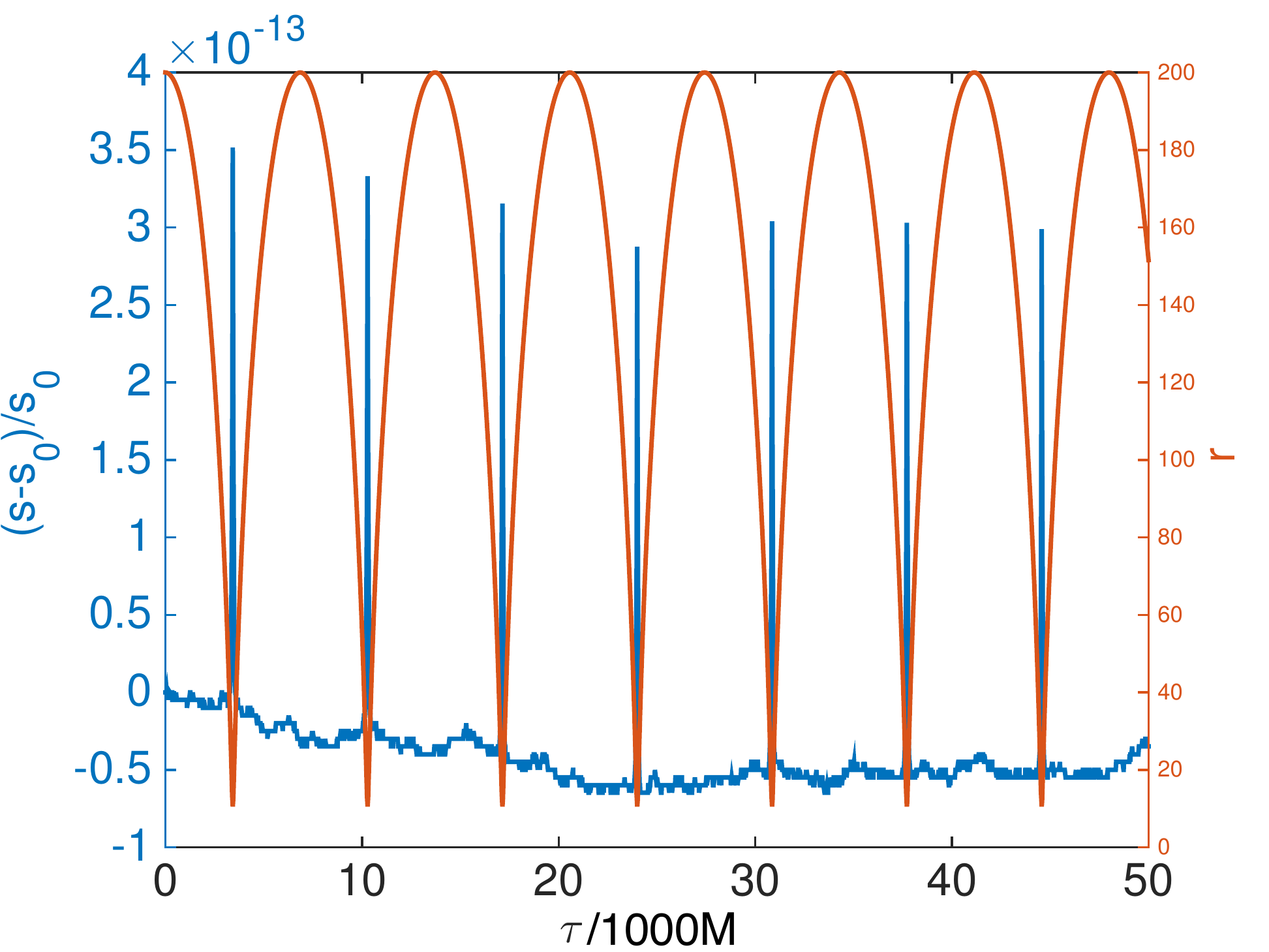}
\caption{Variations of the rotational velocity and orbital radius along evolution time. Top-left panel: $p = 8 M$, $e = 0.2$, $S=10^{-4}$ and $C_Q = 6$; top-right panel: $p = 100 M$, $e = 0.2$, $S=10^{-4}$ and $C_Q = 6$; middle panels: $p = 10 M$, $e = 0.5$, $C_Q = 6$ $S=10^{-4}$ (left) and $S=10^{-5}$ (right); bottom panels: $p = 20 M$, $e = 0.9$, $C_Q = 6$ $S=10^{-4}$ (left) and $S=10^{-5}$ (right).}  \label{fig5}
\end{center}
\end{figure}

Considering the total spin length $S$ is a constant, and $S$ has the dimension of angular momentum. Assuming the radius of body as a constant, $s \equiv S/m$ is proportional to the rotating angular velocity of the extended body.  It means that the rotating velocity is going to its maximum value when the particle is approaching the perihelion. In Fig.\ref{fig5}, we draw temporal variations of $s$ ($s_0$ in this figure means initial value of $s$) and $r$ for the particles with different semi-latus rectums $p$ and eccentricities $e$ in the cases of physical spin magnitude. We find that for the same semi-latus rectums, the orbit with larger eccentricity will produce stronger change of rotational velocity. Obviously, the temporal change of $m$ or $s$ is also directly positive correlation with the value of $C_Q$. This point is too intuitive to need figures for demonstration. 

This mechanism probably makes sense in the pulsar timing observations. The secular stability of some milli-second pulsars is better than atomic clocks on the Earth. The measurement accuracy for the rotational period of pulsars nowadays is incredible good, and the (one sigma) uncertainty can be less than $10^{-15}$ for the certain pulsars \cite{pulsar08}. If the extended bodies are pulsars, the variation of spin $S/m$ in Fig. \ref{fig5} with different orbital parameters in principle can be measured from the pulsar timing observations. Let's see the corresponding astrophysical systems in Fig. \ref{fig5} . We assume a milli-second pulsar with spin period just $10^{-3}$ s and mass $1.4 ~M_{\rm sun}$, then 
\begin{align*} 
\frac{S}{mM} \simeq \frac{2}{5} \frac{R^2}{M} \omega_{\rm NS} \simeq \frac{2}{5} (6.67)^2 M_{\rm sun}^2 \frac{2\pi}{10^{-3} \times 2\times10^5 M_{\rm sun} M} \sim 10^{-4},
\end{align*}
if the mass of central black hole is about $5000 \sim 10^4 M_{\rm sun}$.  The orbital periods are as short as a few minutes. In this estimation, we take the radius of pulsar as 10 km. If $M$ is less than 1000 $M_{\rm sun}$, then $S$ of the orbiting pulsar can achieve $10^{-3}$ order. The mass-ratios of these systems are around $10^{-3} \sim 10^{-4}$, therefore the MPD equations still work though the gravitational self-force corrections are not under consideration. 

Unfortunately, until now we have not yet found such kind of pulsar systems with so small mass-ratios and so short orbital periods.  The above analysis is theoretical and idealized. We hope that the upcoming FAST telescope and SKA will find this kind pulsar-massive blackhole binaries \cite{shao14}. The mechanism of periodic variation in $s$ revealed in this paper may be used to test gravitational theory and constrain the parameters of the pulsars. For a wonderful review of testing relativistic gravity with radio pulsars, please see \cite{pulsar14}.

\subsection{Complex 3D orbits}
For the 3D orbits, the situation becomes very complicated. We can not set in advance the orbital configurations such as $p,~e$ and maximum inclination $\vartheta$ to solve the initial data. As we have shown in Fig. \ref{fig4}, the quadrupole strongly changes the trajectory. The two trajectories and orbital shapes are totally different just after a short evolution.  It is easy to understand such a phenomenon, once $C_Q \neq 0$, the complexity and nonlinearity of MPD equations (\ref{mpd1},\ref{mpd2}) are increased greatly due to the coupling of quadrupole and curvature comparing to the pole-dipole cases.

\begin{figure}
\begin{center}
\includegraphics[height=2.0in]{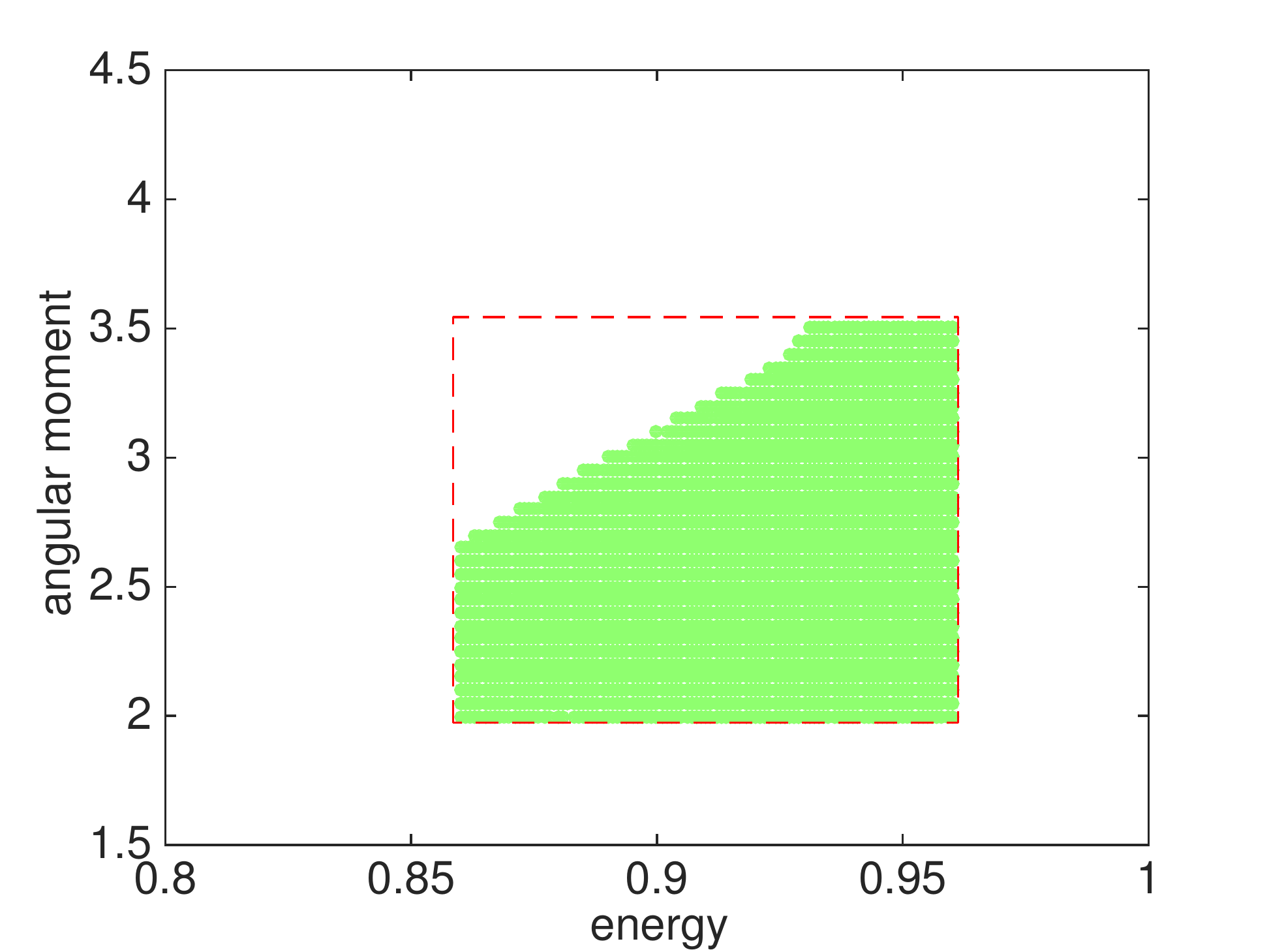}
\includegraphics[height=2.0in]{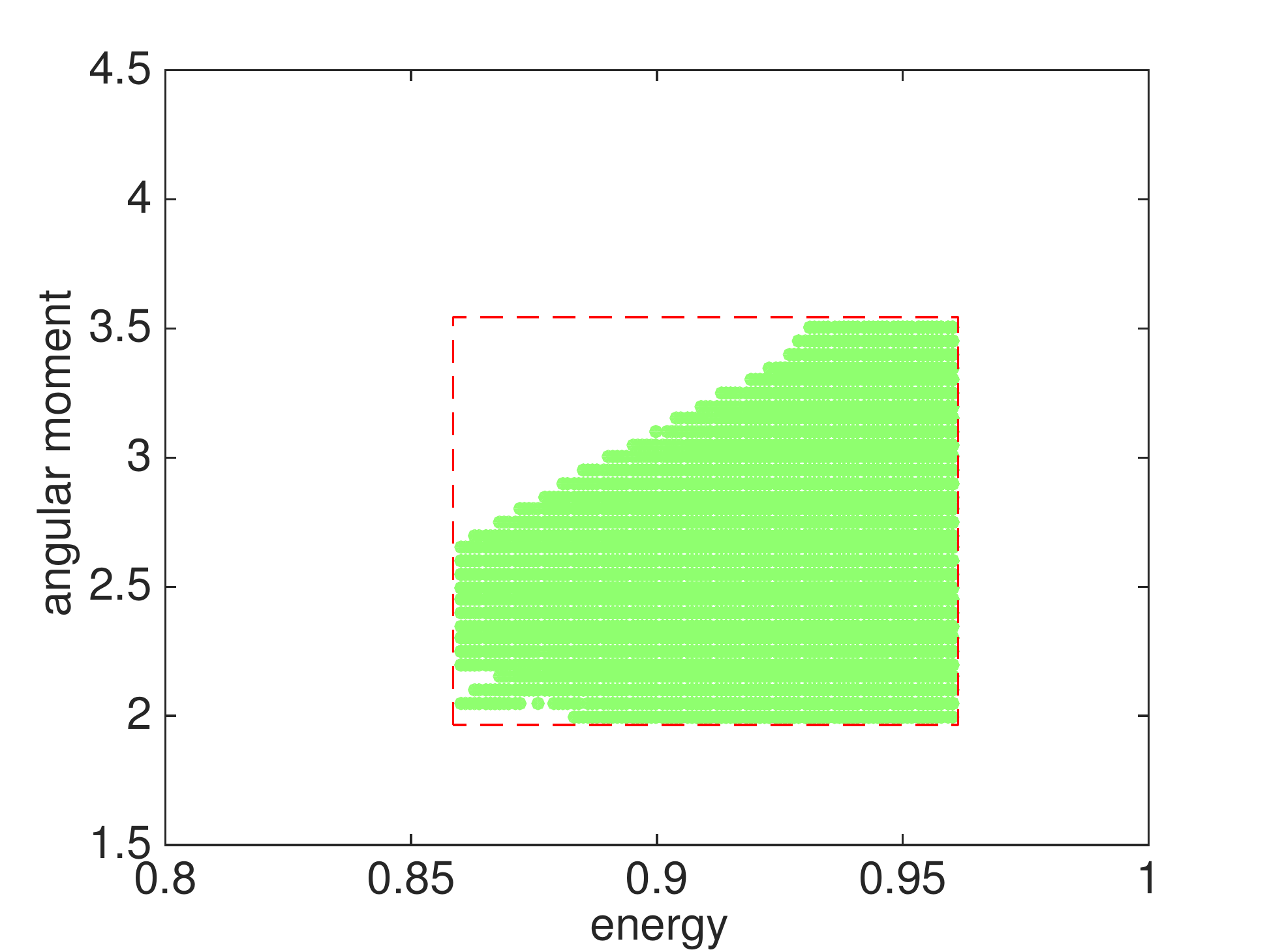}
\includegraphics[height=2.0in]{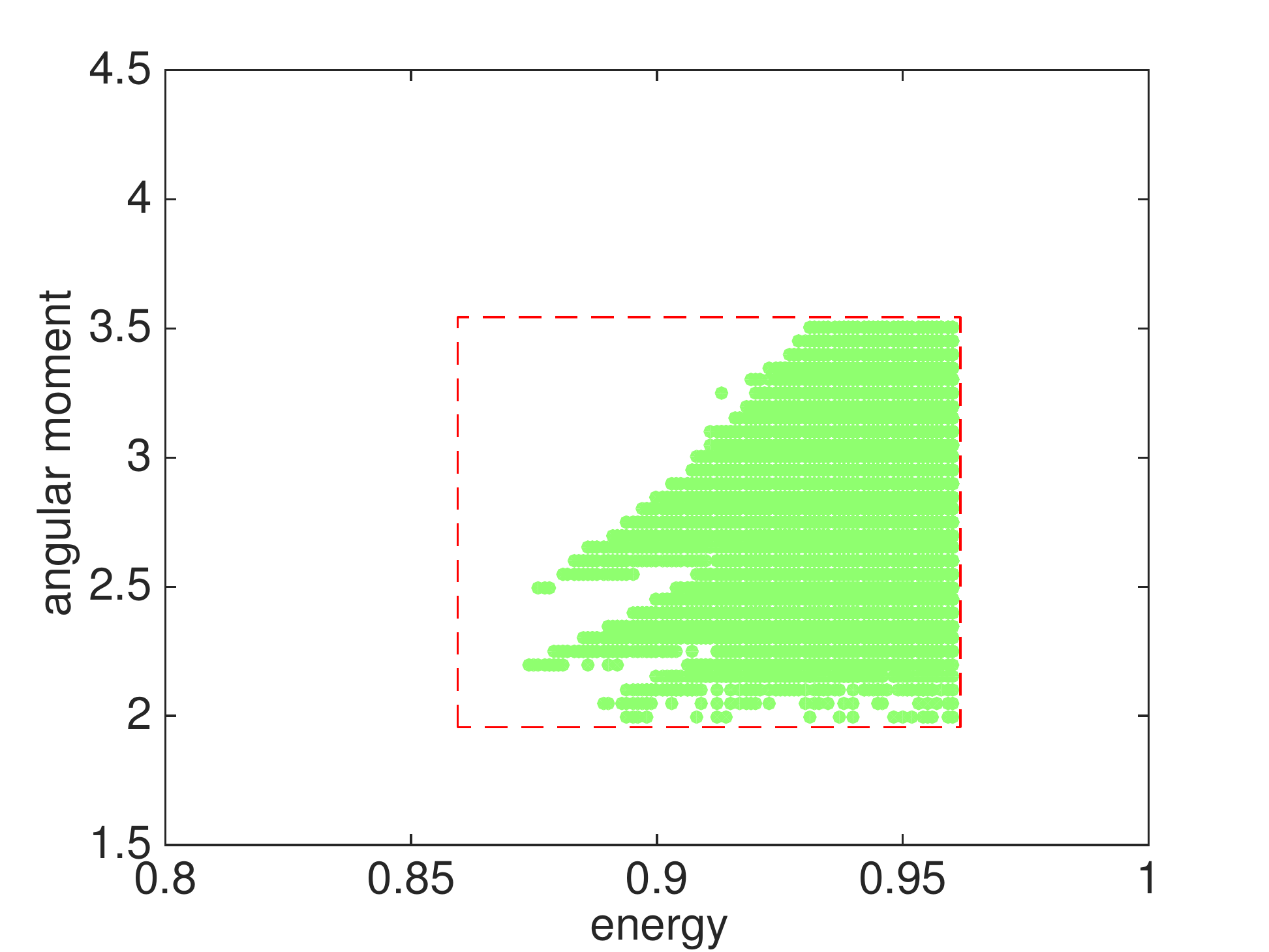}
\includegraphics[height=2.0in]{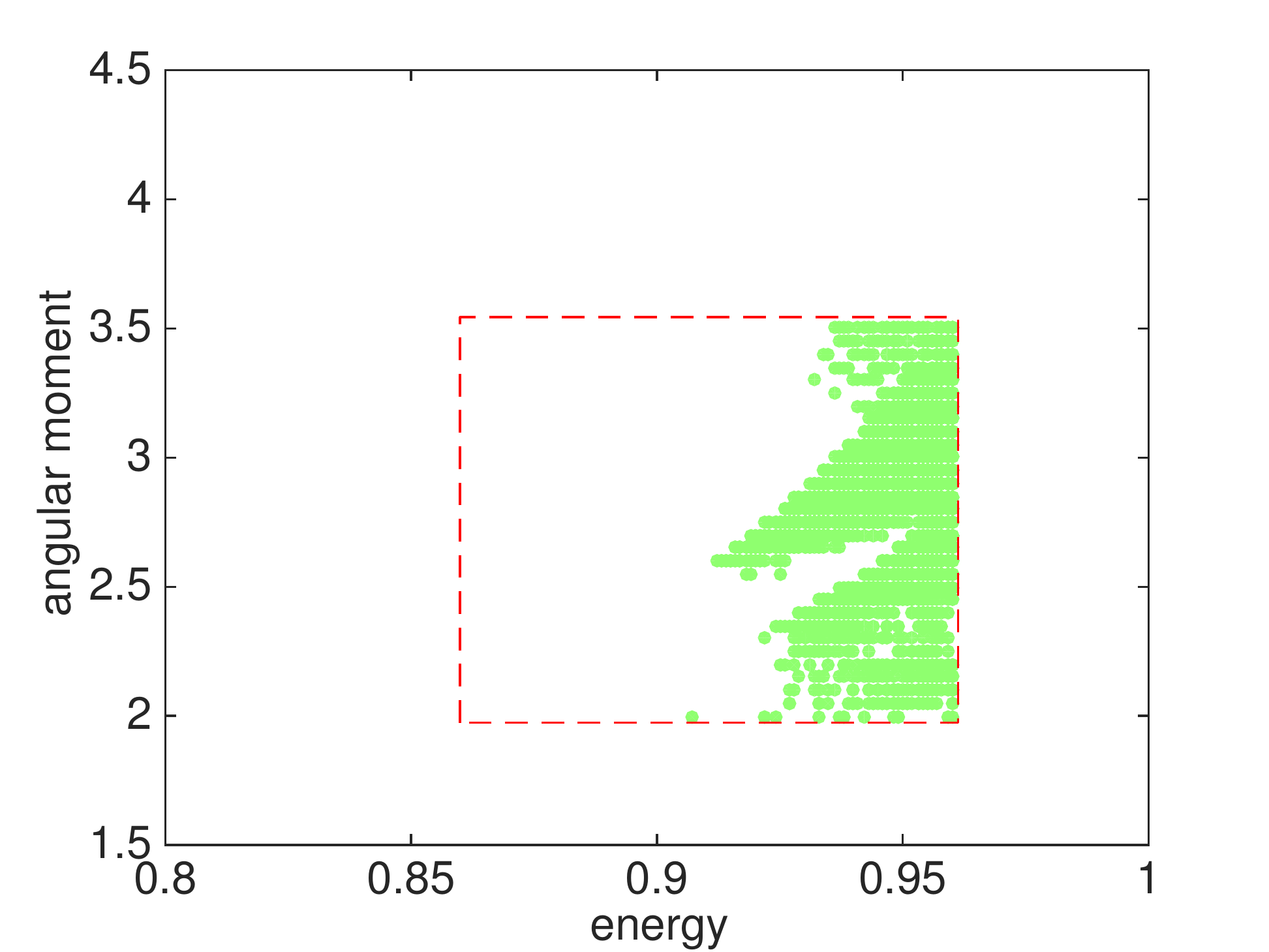}
\caption{Stable orbit zone of spinning extended bodies with $S = 1$ orbiting a Kerr black hole with $a = 1$. The red rectangle is the scanning area, and the green points label the stable orbits. From left to right, top to bottom, $C_Q = $ 0, 1, 4, 8 respectively.}  \label{fig6}
\end{center}
\end{figure}

There are too many issues can be studied  for revealing the effects of the quadrupole. In the present paper, we focus on two issues:  orbital stability and chaos. For the first one, we find that the stable orbits of extremely spinning particles may become unstable after changing $C_Q = 0$ to a nonzero value. With a larger $C_Q$, the possibility of orbital destabilization also goes larger. We scan the parameters $E$ from $0.86$ to 0.96, and $L_z$ from 2.0 to 3.5 with variable $S^r_0, ~S^\theta _0$ and $r_0$. The scanning results shown in Fig. \ref{fig6} clearly demonstrate the decrease of stable orbit zone due to the quadrupole parameter $C_Q$. From the bottom-right panel of Fig. \ref{fig6}, we can find that in the zone with low energy and angular moment, for bodies with $S = 1$ and $C_Q =8$ (maximum estimation of neutron stars), no stably bounded orbit  is found. The orbits with lower energy usually means closer to the central black hole. In other words, neutron stars with larger quadrupoles, can not stably orbit the gravitational center as close as the spinning particles or spinning black holes (which $C_Q = 1$) with the assumption of unphysical spin values. 

However, the obvious influence of the quadrupole on the orbital dynamics can only happen in the case of unphysical spin values. As
we have shown in the section III, because we currently only consider the spin-induced quadrupole moment, the quadrupole effect is quite small when the spin takes physical value ( $< 10^{-2}$) for the extreme-mass-ratio binaries. 

The other interesting issue is the chaos. For the test particles, there are definitely no chaotic orbits because of the integrability of the geodesic equations in the Kerr spacetime. However, a lot of authors have shown that the orbits of extremely spinning particles in the Schwarzschild or Kerr spacetime can be chaotic (see \cite{hartl03, hartl04, han08} and references inside).  One should attend that until now no researchers found chaos for a physical spin value. After many deep searching, Hartl concluded that he find virtually no chaos for spin values below $S = 0.1$. Now, when taking into account the quadrupoles, we will see if or not the quadrupoles have influence on the chaos. Especially, if or not the quadrupoles can make chaos happens in the case of physical spin values. Theoretically, $C_Q \neq 0$ makes the dynamical system more complex and nonlinear, and will be helpful for chaos happening. 

\begin{figure}
\begin{center}
\includegraphics[height=2.0in]{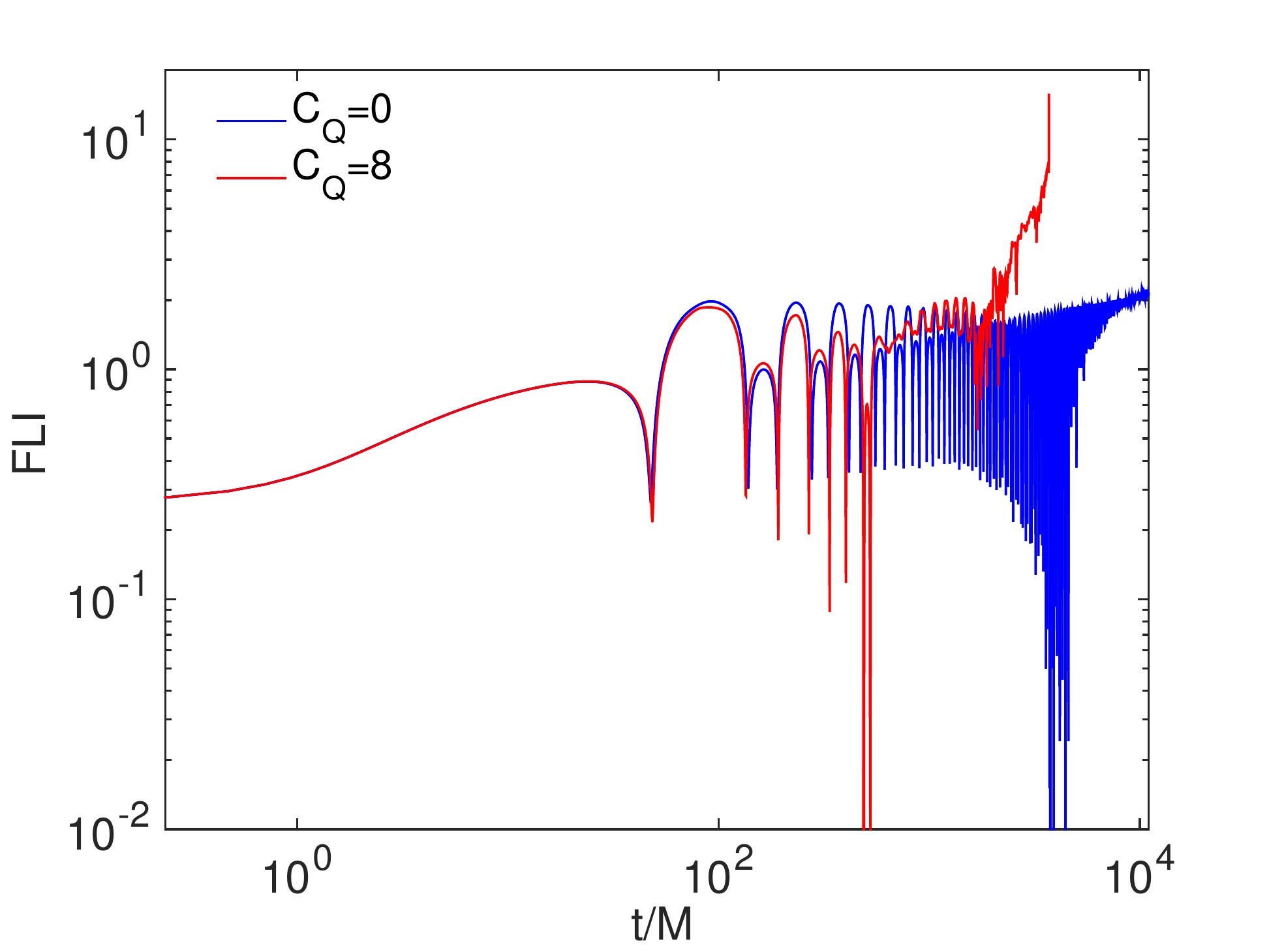}
\includegraphics[height=2.0in]{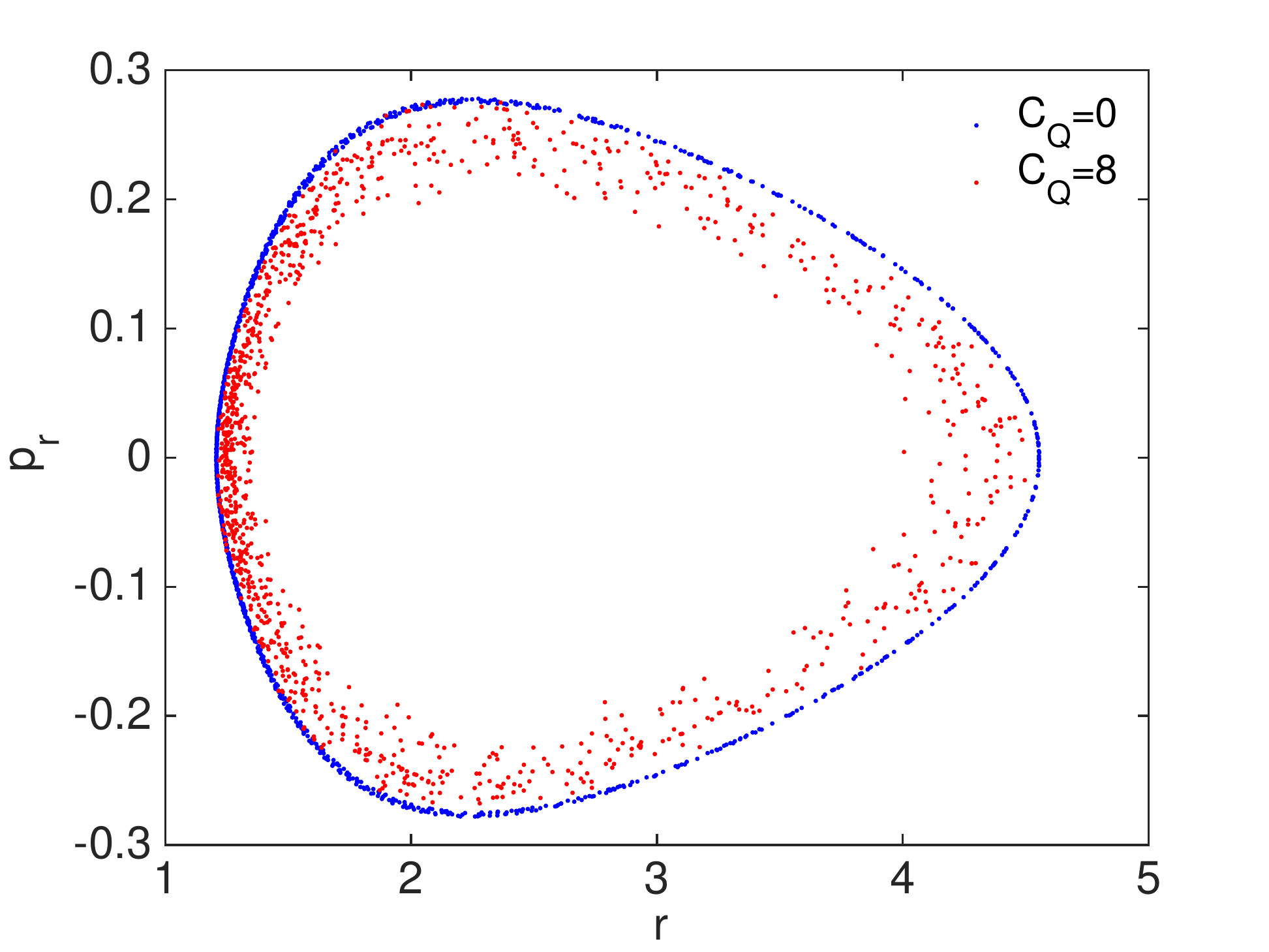}
\includegraphics[height=2.0in]{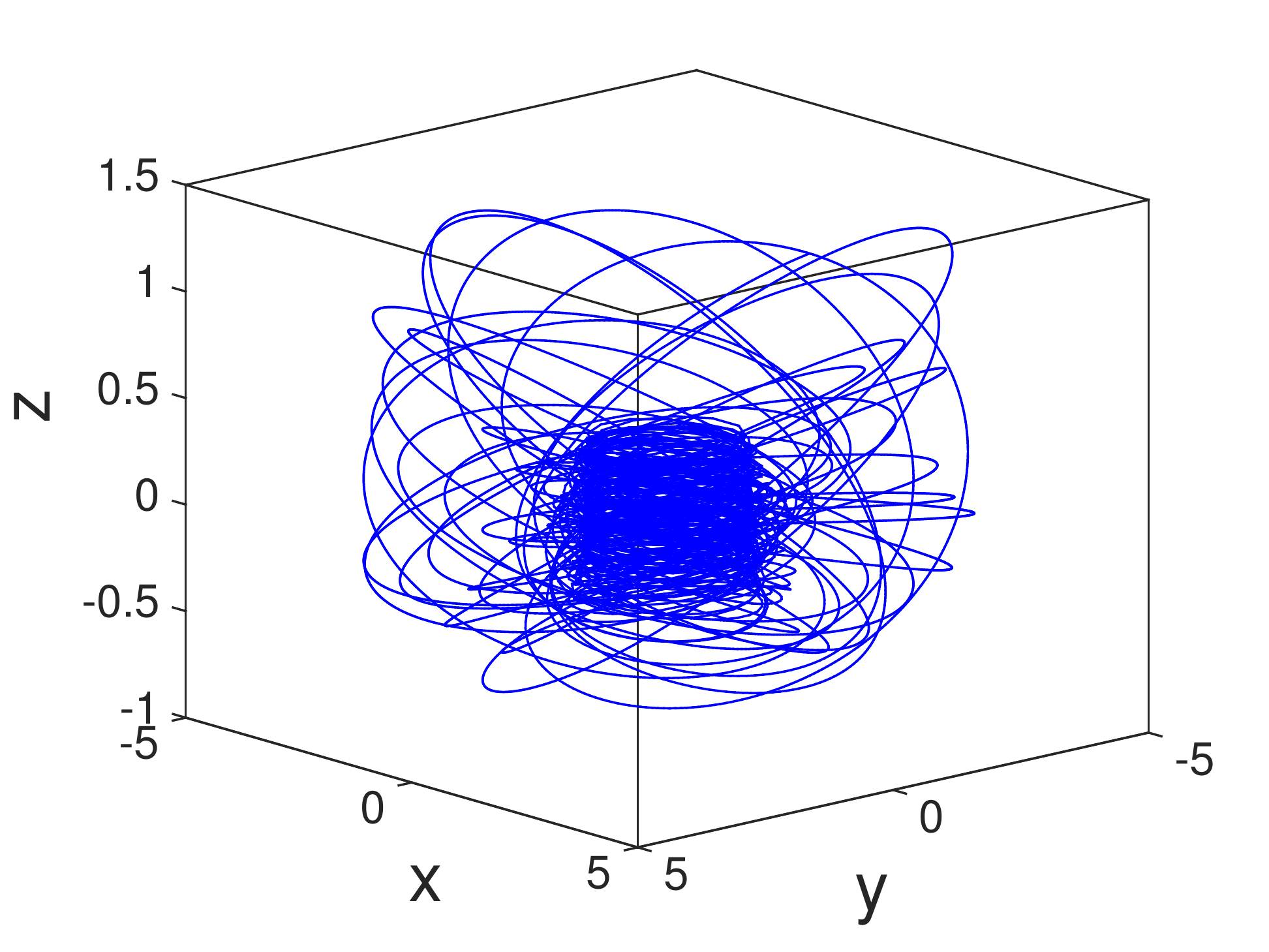}
\includegraphics[height=2.0in]{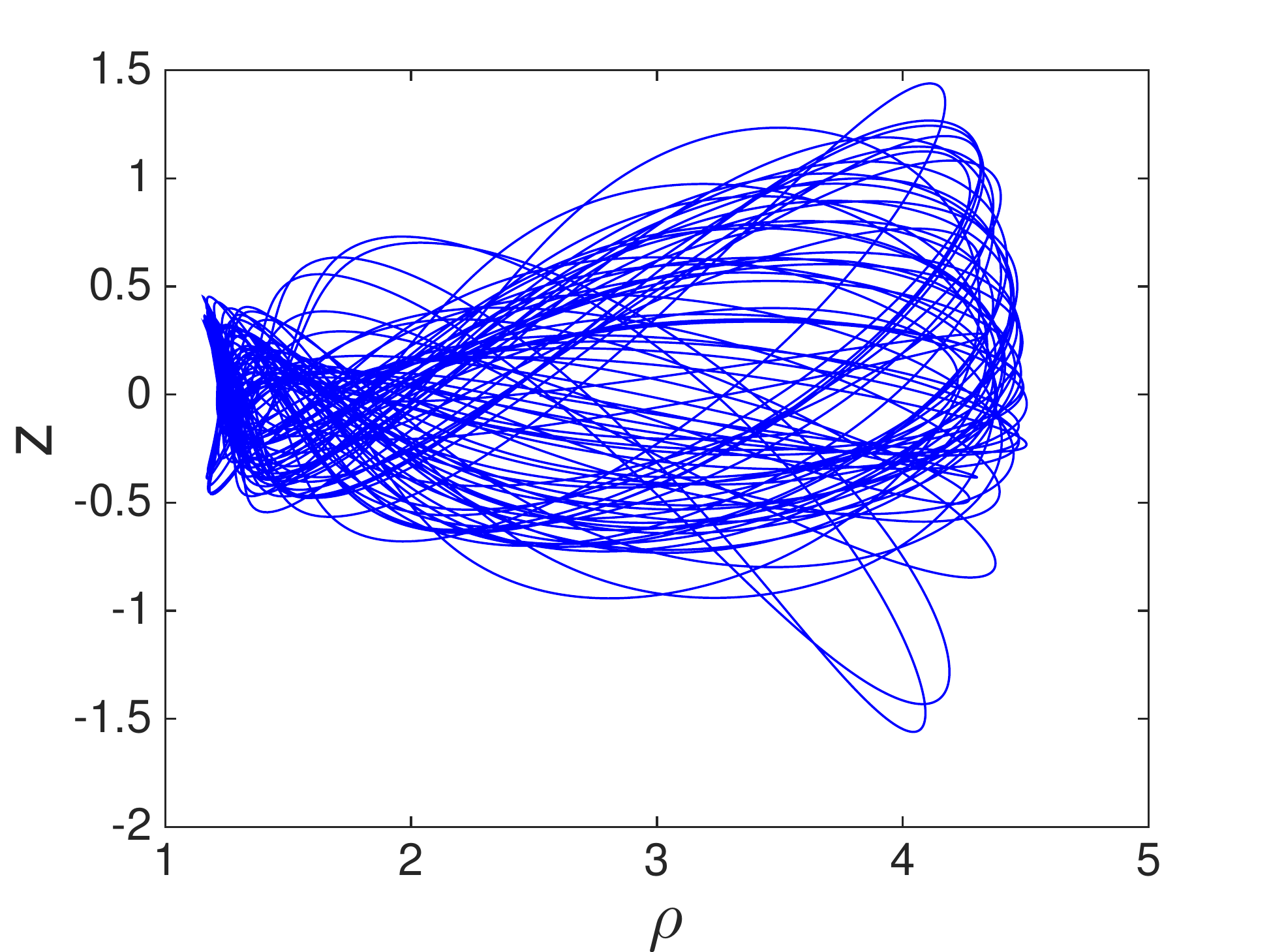}
\caption{Top-left: FLIs of the chaotic orbit (red) and regular one (blue); Top-right: Poinca\'re sections; bottom: trajectory of chaotic orbit. The Kerr black hole has $a=1$, the small body: $S=0.09$, $E=0.84$ and $L_z = 1.75$. For the chaotic case, $C_Q=8$, and initial spatial coordinates are (2.5, $\pi/2$, 0), initial momentum (2.3802, 0.2657, 0.04, 0.4531) and initial spin vector (0.0952, 0, 0, 0.0458). The orbit has $p \approx 1.91, e \approx 0.58$.}  \label{fig7}
\end{center}
\end{figure}

The tool we adopted for fast searching parameter space to find the chaotic orbits is the fast Lyapulov index (FLI) which was firstly introduced by Froeschl\'e et al. \cite{fli1,fli2}. The advantage of FLI is that it need not to be computed until to convergence. Wu et al. introduced the calculation of FLI in curved spacetime \cite{wu06}. We have successfully used FLI to detect chaos in a previous work \cite{han08}. For spinning test particles, no one found chaos when $S<0.1$. However, in the case of extended bodies with quadrupoles, we easily find a lot of chaotic orbits when spin $S=0.09$ and $C_Q=8$. In the top-left panel of Fig. \ref{fig7}, we demonstrate the different behaviors of FLI in chaotic and non-chaotic orbits with $S = 0.09$. After an enough evolution, FLIs of chaotic orbits increase very fast (exponential) in contrast with the regular ones. We also draw the Poincar\'e sections for the both two orbits. It is clearly to find that the quadrupole breaks the regular orbit to a chaotic one. 

\begin{figure}
\begin{center}
\includegraphics[height=2.0in]{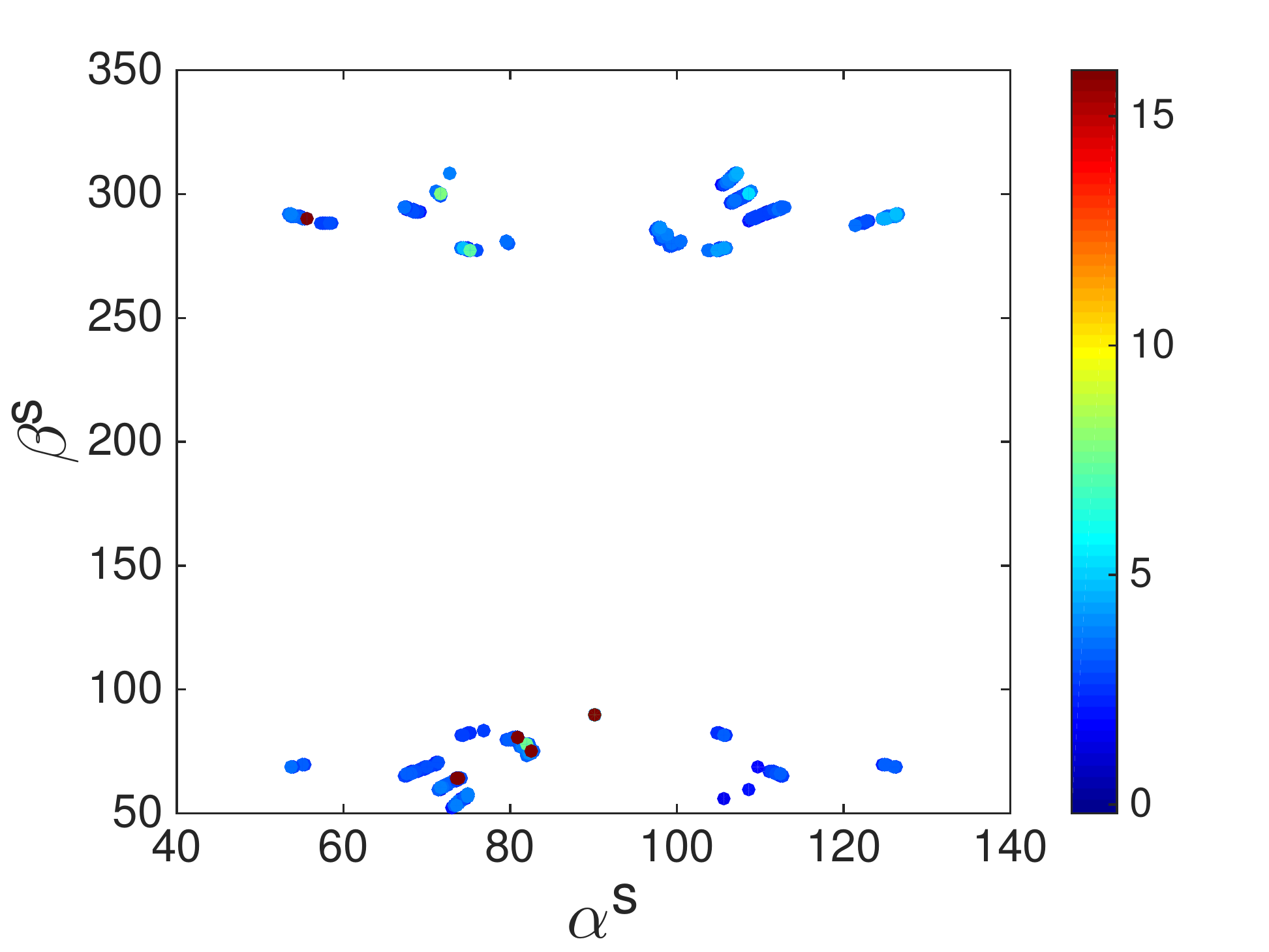}
\includegraphics[height=2.0in]{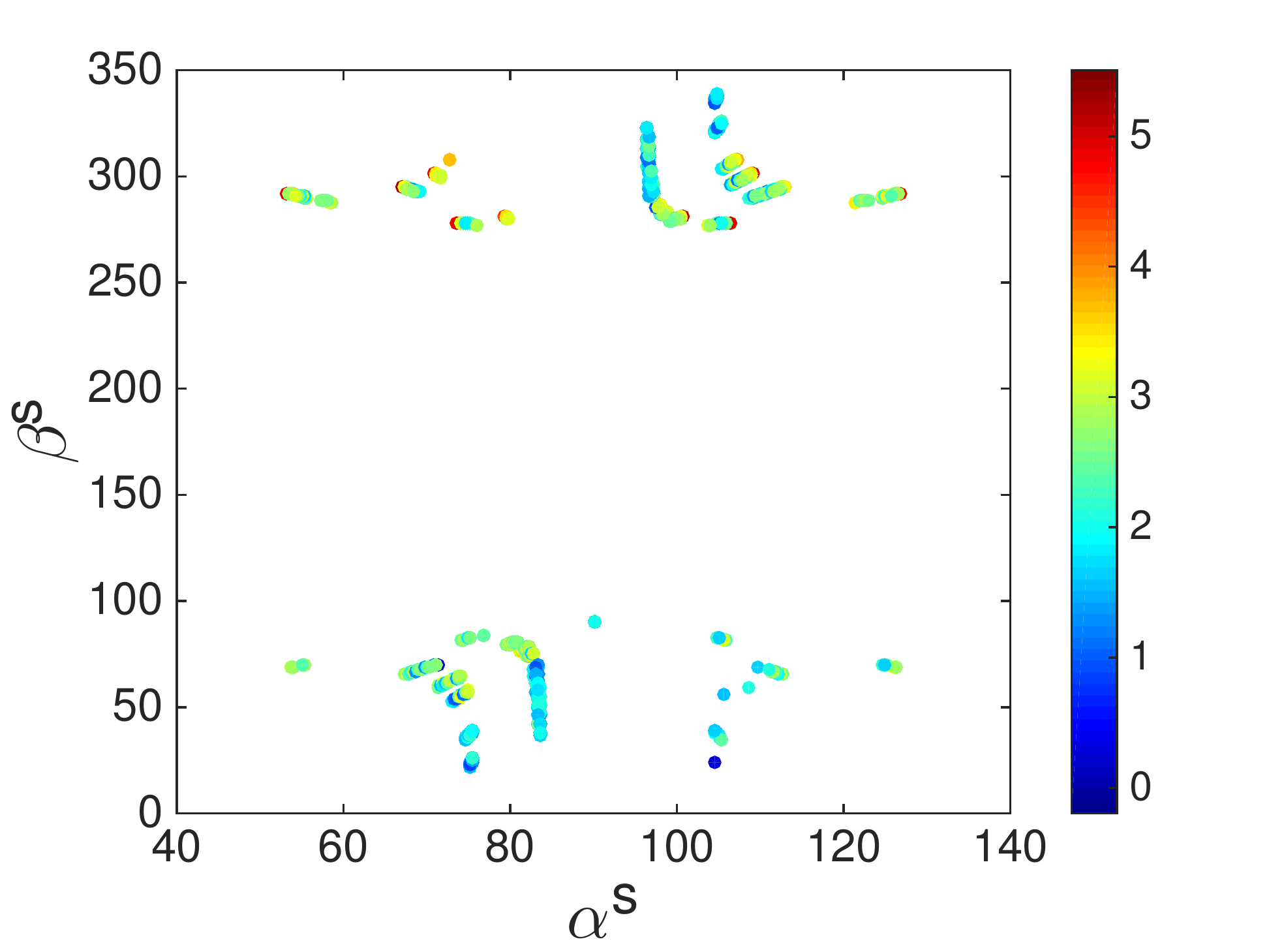}
\caption{Scanning results of FLI for $S=0.05$ (left panel) and $S=0.01$ (right panel) with the quadrupole parameter $C_Q = 8$.}  \label{fig8}
\end{center}
\end{figure}

 In \cite{wu13}, they used a critical value of FLI to identify the chaos and regularity. In this paper we decide to adopt the same critical value we used in \cite{han08}, i.e.,  at the end of orbital evolution ($t$ reaches $5\times10^4 M$), if FLI is less than 6, we conclude that the orbit is regular (be careful that this is not a very rigorous identification). In Fig. \ref{fig8}, values of FLIs of variable initial angles of spin vectors are plotted. The orientations of spin vectors are presented in the observer's local orthonormal space triad, see \cite{semerak99} for details. The scanning results tell us there are still a few of chaotic orbits (FLIs $> 6$) even for $S=0.05$ with $C_Q = 8$. However, when $S$ reduces to 0.01, no chaotic orbit can be found (FLIs $<6$). So, we can cautiously conclude that no chaos can happen when $S < 0.01$ even for $C_Q = 8$ (the maximal estimation for neutron star). In one word, we have found chaos in the cases of $0.01 < S < 0.1$ for spin-induced deformed bodies, previously no chaos was found for spinning test particles with $S < 0.1$. 

Therefore, we believe that there is no chaotic orbits for the realistic extreme-mass-ratio binaries.  However, this conclusion valid only for the cases in which the quadrupoles are spin-induced. It should be very interesting to include the tidal-induced quadrupoles in the next work. However, with spin-induced one, we have succeeded finding chaos for smaller spin magnitude than the previous literature. 

\section{conclusions and discussions}
Large mass-ratio compact binaries are very important sources for the low frequency GW detectors. Dynamical model without enough accuracy for systems of such kind will cause serious dephase of the GW templates and induce a failure of finding GW signals. Especially, GWs from chaotic orbits can not be detected because it is hard to be predicted and modeled accurately. At the same time, such systems which includes pulsars are interested to pulsar timing observations and can be important tools to probe the strong field gravity and test general relativity. 

In the present paper, we derived an closed form of the relation of four velocity and momentum from the orthogonal condition $\upsilon^\mu u_\mu = -1$.  Nevertheless, our relation in principle has not difference with the one given by \cite{ehlers77}.  
We carefully analyze the effect of spin-induced quadrupoles on the circular orbits, and we find that for the physically allowed spin values of EMRIs, such kind of quadrupoles can be safely omitted in the GW modeling. For eccentric-equatorial orbits, we reveal the variation of dynamical mass vs. the orbital radius. We then expect that the periodic variation of rotation of pulsars in such kind systems can be measured in the future pulsar timing observations. 

Finally, for the complex 3D orbits, with the help of numerical integrations, we find that the quadrupoles will reduce stable zone of orbits in the case of artificial large spin. This is because the $C_Q$ greatly enlarge the complexity and nonlinearity of the dynamical systems. Based on the same mechanism, the quadrupoles will also promote the appearance of chaos when $S < 0.1$. Before this work, there is no report of finding chaos for the spinning particles when spin is less than 0.1. However, we still did not find chaos for much smaller $S$ ($< 0.01$) if the quadrupoles are produced by the spins only. This implies that chaotic orbits may not appear in the astrophysical EMRIs. 

In one word, spin-induced quadrupoles may not have any influences on detecting GWs from EMRIs, but do have considerable effects for pulsar timing observations of the compact pulsar-massive black hole binaries. The later one can be used to constrain the equation of state of neutron star and probe the strong gravity. We will continue working on this issue by considering tidal-induced and generic quadrupoles in the next paper.

\section*{Acknowledgement} 
 WH appreciates Prof. Kopeikin for useful discussions. This work is supported by NSFC No. U1431120, No.11273045, and QYZDB-SSW-SYS016 of CAS. WH is also supported by Youth Innovation Promotion Association CAS. 
\nocite{*}

\bibliography{apssamp}
{}    
\end{document}